\documentclass[twocolumn,superscriptaddress,amsmath,floatfix,amssymb]{revtex4}

\usepackage[colorlinks=false]{hyperref}
\usepackage{graphicx}

\newcommand{\beqn}{\begin{eqnarray}}
\newcommand{\eeqn}{\end{eqnarray}}
\newcommand{\eq}[1]{(\ref{#1})}

\newcommand{\Gc}{\text{\fontsize{10pt}{14pt}Hz}}
\newcommand{\Hz}{\text{\fontsize{10pt}{14pt}Hz}}

\newcommand{\bohr}{\text{\fontsize{10pt}{14pt}bohr}}
 
\newcommand{\Cc}{\text{\fontsize{10pt}{14pt}$^o$C}}

\newcommand{\Kev}{\text{\fontsize{10pt}{14pt}keV}}
\newcommand{\Mev}{\text{\fontsize{10pt}{14pt}MeV}}

\newcommand{\MGc}{\text{\fontsize{10pt}{14pt}MHz}}
\newcommand{\ccm}{\text{\fontsize{10pt}{14pt}cm}}
\newcommand{\cm}{\text{\fontsize{10pt}{14pt}cm}}
\newcommand{\m}{\text{\fontsize{10pt}{14pt}m}}
\newcommand{\nm}{\text{\fontsize{10pt}{14pt}nm}}
\newcommand{\mm}{\text{\fontsize{10pt}{14pt}mm}}
\newcommand{\cc}{\text{\fontsize{10pt}{14pt}s}}
\newcommand{\ns}{\text{\fontsize{10pt}{14pt}ns}}
\newcommand{\mcc}{\text{\fontsize{10pt}{14pt}ms}}

\newcommand{\ncc}{\text{\fontsize{10pt}{14pt}ns}}
\newcommand{\Dg}{\text{\fontsize{10pt}{14pt}J}}
\newcommand{\mDg}{\text{\fontsize{10pt}{14pt}mJ}}

\newcommand{\Wt}{\text{\fontsize{10pt}{14pt}W}}
\newcommand{\mWt}{\text{\fontsize{10pt}{14pt}mW}}
\newcommand{\W}{\text{\fontsize{10pt}{14pt}W}}
\newcommand{\mW}{\text{\fontsize{10pt}{14pt}mW}}
\newcommand{\kWt}{\text{\fontsize{10pt}{14pt}kW}}
\newcommand{\nWt}{\text{\fontsize{10pt}{14pt}nW}}
\newcommand{\Tl}{\text{\fontsize{10pt}{14pt}T}}
\newcommand{\T}{\text{\fontsize{10pt}{14pt}T}}
\newcommand{\K}{\text{\fontsize{10pt}{14pt}K}}
\newcommand{\mK}{\text{\fontsize{10pt}{14pt}mK}}

\newcommand{\nK}{\text{\fontsize{10pt}{14pt}nK}}
\newcommand{\Dop}{\text{\fontsize{10pt}{14pt}dop}}
\newcommand{\cool}{\text{\fontsize{10pt}{14pt}cool}}
\newcommand{\heat}{\text{\fontsize{10pt}{14pt}heat}}
\newcommand{\mmin}{\text{\fontsize{10pt}{14pt}min}}
\newcommand{\kg}{\text{\fontsize{10pt}{14pt}kg}}
\newcommand{\rec}{\text{\fontsize{10pt}{14pt}rec}}
\newcommand{\dip}{\text{\fontsize{10pt}{14pt}dip}}
\newcommand{\st}{\text{\fontsize{10pt}{14pt}rest}}

\newcommand{\opt}{\text{\fontsize{10pt}{14pt}opt}}
\newcommand{\rad}{\text{\fontsize{10pt}{14pt}rad}}
\newcommand{\pK}{\text{\fontsize{10pt}{14pt}pK}}
\newcommand{\micr}{\text{\fontsize{10pt}{14pt}micr}}

\begin{document}

\sloppy

\title{Current progress in laser cooling of antihydrogen}

\author{E.\,V.\,Luschevskaya}
  \email{luschevskaya@itep.ru}
\author{A.\,A.\,Golubev$^{2,}$}%
 \email{alexander.golubev@itep.ru}
\affiliation{
National research centre "Kurchatov Institute"  Institute of Theoretical and Experimental Physics, Bolshaya Cheremushkinskaya 25, Moscow, 117218, Russia\\
$^2$National Research Nuclear University,  Kashirskoye shosse 31, Moscow, 115409, Russia% with \\
}%

\begin{abstract}
We discuss laser cooling methods  of (anti)hydrogen  and its importance for  current and future experiments.
The exploration of antimatter presents a great interest for $CERN$ and $GSI$ experiments aimed at  check of quantum mechanics laws, fundamental symmetries of nature and gravity and investigations in atomic and nuclear physics.
The spectral transition $1S\rightarrow 2P$ in  $\bar{H} (H)$  atom   is the most suitable for laser cooling due to  a  small lifetime of $2P$ state and insignificant  ionization losses.  However the   pulsed and continuous laser sources at Lyman-$\alpha$ wavelength do not possess enough power for fast and efficient cooling. The small power of laser sources at $\lambda=121.6\ \nm$   is poor technical problem associated with a complexity  of generation scheme of such radiation, which arises due to absence of nonlinear $BBO$ crystals  at this wavelength.   The advances in this area will completely destine the future progress of the experiments aimed at study of antimatter. 
 \end{abstract}

\maketitle

\section{Introduction}

\label{intro}

In 1931 Paul Dirac predicted the existence of  positron  \cite{Dirac:1928,Dirac:1931}, which
was discovered  in cosmic rays by Karl Anderson two years later \cite{Anderson:1932,Anderson:1933}. In 1955  antiproton  was found on the accelerator at the University of California.

In accordance with the $CPT$ theorem any Lorentz invariant local quantum field theory with  hermitian Hamiltonian must obey
the $CPT$ symmetry \cite{Schwinger:1951}. In other words there is  antiparticle for every particle with the same mass, spin and  lifetime, but with the  opposite charge and magnetic moment. Violations of this theorem would imply  new physical properties of  fields and their interactions.

Antiparticles can be combined into antiatoms and antimolecules.  An  antihydrogen atom $\bar{H}$, the  bound state of  antiproton and  positron, represents the simplest element of antimatter.
In 1995 at $CERN$ antiproton ring $LEAR$  the first evidence of the antihydrogen existence was obtained.
 Nine antiatoms were detected, which moved with relativistic energies.
It was impossible to make physical measurements because the observation period of antiatoms was very small \cite{Baur:1996}.

 In 1996 at Fermilab several dozens of antiatoms were found in a similar experiment  \cite{Blanford:1998}.  It is worth noting  that  antihelium nuclei were detected at $IHEP$ in 1969  \cite{Antipov:1970}, much earlier than antihydrogen at $CERN$. 

The construction of  $CERN$ Antiproton Decelerator (AD) started in 1999 \cite{Maury:1997}. Currently this machine decelerates antiprotons down to energy  $\sim 5.3\ \Mev$, it delivers $\sim 3\times 10^7\ \bar{p}$ per pulse $\sim 150\ \ncc$, the duration of the cycle is $110\ \cc$. Then antiprotons  move to various experiments. $ATRAP$ \cite{ATRAP} and $ALPHA$ \cite{ALPHA} experiments catch the antiprotons in a Malberg-Penning trap for further deceleration and  cooling.
$ASACUSA$ \cite{ASACUSA}, $ACE$ \cite{ACE}, $AE\bar{g}IS$ \cite{AEGIS}, $BASE$ \cite{BASE,BASE2} experiments  use a beam of antihydrogen atoms.     $GBAR$ collaboration also  plans to use a beam of antiatoms for further investigations \cite{GBAR}.
 The   Extra Low ENergy Antiproton ring was proposed $(ELENA)$ \cite{ELENA} to increase the number of accumulated antiprotons by one or two orders of magnitudes. $ELENA$  will become  an upgrade of the Antiproton Decelerator. This new machine will supply   $CERN$ experiments with antiprotons of $100\ \Kev$ energy. So low energies could be achieved due to electron cooling   \cite{Menshikov:2008}.

For synthesis of  antihydrogen atoms   antiprotons are combined  with positrons accumulated preliminarily in a positron trap  \cite{Surko:1992,Surko:2003,Surko:2004}.
$ALPHA$ collaboration obtains and  detects   antihydrogen since 2002 at $CERN$ \cite{Amoretti:2002}, but at first they localized the antiatoms within the magnetic trap  only in 2010 during $172\ \mcc$ \cite{Andresen:2010}. In 2011 confinement time of  antihydrogen was  approximately equal to $1000\ \cc$ \cite{Andresen:2011,Amole:2012}.
 In 2012 $ATRAP$ collaboration also synthesized  the antihydrogen atoms in the magnetic trap \cite{Gabrielse:2012}.

 At  current experimental facilities  antihydrogen formation occurs  due to three-body recombination \cite{Hess:1984,Gabrielse:1988,Glinsky:1991,Robicheaux:2004}
 \begin{equation}
 \bar{p}+e^+ + e^+ \rightarrow \bar{H} + e^+
 \end{equation}
   and recombination via   resonant charge exchange \cite{Hessels:1998,Storry:2004} 
    \begin{equation}
 \bar{p}+ Ps^* \rightarrow \bar{H}^* + e^+,
 \end{equation}
 where $Ps^*$ is the Rydberg positronium,  the bound state of electron and positron in the excited state which can be produced via charge exchange collisions \cite{Speck:2004}, $\bar{H}^*$ is the antihydrogen atom in an excited state.
       Theoretically $\bar{H}$ can be also produced via pulsed field recombination \cite{Wesdorp:2001}, 
radiative recombination \eqref{eq01} and laser-induced recombination  \eqref{eq02} \cite{Gabrielse:1988}
  \begin{equation}
\bar{p}+e^+ \rightarrow \bar{H} + h\nu,
\label{eq01}
 \end{equation}
 
\vspace{-0.4cm} 
 
 \begin{equation}
 \bar{p}+Nh\nu\rightarrow \bar{H}+ (N+1)h\nu.
 \label{eq02}
 \end{equation}
$CERN$ collaborations  synthesize  antihydrogen for  various goals, such as a search of distinctions between hydrogen and antihydrogen, gravitaional experiments, study of antiprotonic helium, measurement and comparison of the magnetic moments of proton and antiproton, antiproton therapy for cancer treatment and nuclear physics.

Theoretical studies show that violations of the $CPT$ symmetry,  if they exist, could be manifested
in  changes of  frequencies of  spectral transitions, Lamb shift and hyperfine structure of antihydrogen in comparison with hydrogen  \cite{Parkhomchuk:1988,Bluhm:1999,Hansch:1993}.
The deviations from the Standard Model can reveal itself in the violations of Lorentz and CPT symmetries.
 Antiatoms  like any other antimatter can be a  tool for testing of  interaction dynamics of atoms and antiatoms   \cite{Campeanu:1983}. The theoretical review about  cold antihydrogen  can be found  in   \cite{Menshikov:2003}.

Collaboration $FLAIR$ (Facility for Low-energy Antiproton and Ion Research) in $GSI$ research center in Darmstaadt    also plans the study of antimatter and antihydrogen physics at their experimental facility.
Their proposal is presented in \cite{Widmann:2005}. It is supposed to improve the  accuracy of precision experiments carried out at $CERN$ owing to a special antiproton slowing-down system.  The main part of $FLAIR$ research aimed at study of antimatter and its interaction with matter and in particular with heavy nuclei. 

The exploration of antimatter is an exciting scientific topic and   undoubtedly deserve a large attention. W. Heisenberg in 1972  
said: "I think that the discovery of antimatter was perhaps the biggest jump in physics in our century" \cite{Heisenberg:1972}.

 \section{History of laser cooling}
 \label{sec:1}

%!!!!!!!!!!!!!!!!!!!!!!!!!!!!!!!!!!!!!!!!!!!!!!!!!!!!!!!!!!!!!!!!!!!!!!!!!%
%(!!!!!!вставить ссылки на другие обзоры, их гораздо больше!!!!!!!!!!!!! )%
%!!!!!!!!!!!!!!!!!!!!!!!!!!!!!!!!!!!!!!!!!!!!!!!!!!!!!!!!!!!!!!!!!!!!!!!!!%
 
In 1975 T. Hansch and A. Schawlow  \cite{Hansch:1975}  and 
independently D. Wineland and H. Dehmelt \cite{Dehmelt:1975} proposed a new cooling  method using laser irradiation 
 based on  Doppler effect. 
Later the first inspiring experiments   were carried out with $Mg^+$ and $Ba^{+}$ ions held in a magnetic trap \cite{Wineland:1978,Neuhauser:1978}.
This  amazing action of light  has many applications in science today.
 A large number of outstanding theoretical and experimental studies have been devoted to  various schemes of laser cooling of neutral atoms, a reader can study the nobel papers  \cite{nobellectures,Phillips:1998nl,Chu:1998nl,Cohen:1998nl}, reviews \cite{Balykin:2011,Balykin:1985,Phillips:1989,Metcalf:2003} and books \cite{book:1,book:2,book:3}.

  In  1976 V.S. Letokhov, V.G. Minogin and B.D. Pavlik proposed the   chirp-cooling technique 
    \cite{Letokhov:1976}. 
According to this method the frequency detuning of the cooling laser  varies in time so that already chilled atoms stay in resonance with the light wave.
This suggestion was  successfully implemented   in the experiment of D. Prodan and W. Phillips in 1983 \cite{Prodan:1983,Prodan:1984}.
 
 Later the theory of Doppler cooling was developed and improved significantly. 
 The lowest temperature which could be achieved in this method was found for various conditions, for example in the presence of external magnetic field  \cite{StenholmS:1978,Javanainen:1980,LetokhovV:1977,JavanainenJ:1980,Cook:1980,Wineland:1979,Stenholm:1986}.

 Zeeman cooling method was developed in 1982 by G. Metcalf  \cite{Metcalf:1982}. Atoms moved from a high field region to a low field one as the atomic velocity decreased. So the shift of Zeeman sublevels compensates   the decrease of Doppler shift and the laser frequency is   maintained constant. 
 Due to Zeeman cooling sodium atoms were slowed down to  the velocity $\backsimeq 40\ \m / \cc$ \cite{Phillips:1982} and soon afterwards to $15\ \m / \cc$  corresponding to the temperature $\sim 100\, \mK$ \cite{Phillips:1985}. The details of these experiments are also presented in \cite{ProdanJ:1982,PhillipsW:1984,MetcalfH:1985,PhillipsW:1985}

 In 1985 the group of scientists under the guidance of A. Migdal caught the sodium atoms in the quadrupole magnetic trap for one second  \cite{Migdall:1985}, the trap depth was $17\, \mK$ which corresponds to the atomic velocity of $3.5$ m/s.

In 1987  W. Phillips and his colleagues found  disagreements between  Doppler cooling theory and their experimental results. They studied the lifetime of optical molasses as a function of the laser frequency detuning. They got the minimal temperature   $43 \pm 20\ \mu K$   much lower than the Doppler limit which is equal to $240\ \mu K$     \cite{Phillips:1988,Phillips:1989}.
 
 J.Dalibard, C.Cohen-Tannoudji and S.Chu  predicted and explained the observed  effects  from the theoretical point of view \cite{Dalibard:1989,Ungar:1989,Dalibard:1985}.
 They showed that in a standing wave  at low  intensity of  laser wave   the subdoppler cooling might be provided either by spatial gradient of  polarization or variable intensity.  
The minimal temperature was  strongly dependent on the magnetic field value.  So the subdoppler cooling was discovered.
In 1997 J.Dalibard, C.Cohen-Tannoudji and S.Chu were awarded  the Nobel Prize for their developments of cooling and trapping methods of atoms by laser field.

For the laser irradiation of high intensity the Sisyphus cooling scheme was suggested   \cite{Wineland:1992}. A strict correspondence between the theory and experiment hadn't been observed, but at the qualitative level the theory was confirmed by  experimental results.
%!!!!!!!!!!!!!!!!!!!!!!!!!!!!!!!!!!!!!!!!!!!!!!!!!!!!!!!!!!!!!!!!!!!!!!!!!%
%(!!!!!!может быть, стоит кратко сказать, какие именно экспериментальные результаты это были?!!!!!!!!!!!!! )%
%!!!!!!!!!!!!!!!!!!!!!!!!!!!!!!!!!!!!!!!!!!!!!!!!!!!!!!!!!!!!!!!!!!!!!!!!!%

In 1990 the first evidences of optical lattices were obtained    \cite{Weiss:1989} and in the same year their first investigation had been carried out  \cite{Westbrook:1990}.
In 1990 W. Phillips with collegues cooled cesium atoms to    $2.5\ \mu K$ \cite{Solomon:1990}.  

The next record temperature was achieved due to the method of adiabatic expansion,  W. Phillips and his collegues reach the temperature  $700\ \nK$  for  sodium atoms \cite{Kastberg:1995}. 
However so low   temperatures can be obtained only for the alkaline metals.
The cases of helium and hydrogen are not so optimistic. 
 There is some recoil temperature  $T_R$ determined by the relation $T_R=2E_R/k_B$, where $E_R$ is the recoil energy, $k_B$ is the Boltzmann constant. 
  For helium the recoil temperature is of the order of $ \mu K$,   for hydrogen atom it lies in the $ \mK$ temperature range. 
This limit may be overcommed owing  to  the well known methods: 
 velocity-selective coherent population trapping    \cite{Aspect:1988}, Raman cooling \cite{Kasevich:1992} and sideband cooling for ions \cite{Dehmelt:1975}. 
In all these technics the spontaneous irradiation processes are   velocity dependent 
to prevent the light scattering by the coldest atoms.
 
  In 1994 the velocity-selective coherent population trapping  allows  to get   the temperature   $T_R/2$ \cite{Bardou:1994} for the beam of helium atoms   and  in 1997 the temperature $T_R/800$ for the precooled helium sample   \cite{Saubamea:1997}.  
 
    If we would have the laser sources required wavelength and power,  these  methods and their modifications may be applied for cooling of (anti)hydrogen.  
A  laser cooling   can be a preliminary procedure  followed  by  evaporative cooling to get more dense and large condensates of atomic hydrogen.  Cooling  below   one-photon  recoil    needs     laser  sources  of  very  high  power.

 \section{Theory of laser cooling}
 \label{sec:2}
 
 \subsection{Radiation pressure} 
\label{subsec:3}

For the first time the assumption about interaction of   light wave  with a matter was made by J. Kepler in 17th century. He  explained why   tails of comets are elongated in the opposite direction from the Sun. In the 19th century Maxwell obtained the formula for the force of radiation pressure. 

In 1901 P. Lebedev, E. Nichols and G. Hull   in laboratory experiments proved that  the light   exerts pressure on various objects \cite{Lebedev:1901:1,Nichols:1901,Nichols:1903}.
 The experiment confirmed the theory despite a lack of a  good vacuum.  In 1910 P. Lebedev obtained more accurate results, which prove the Maxwell theory of light pressure \cite{Lebedev:1901:2}.
  
Recoil phenomena  arising due to reflection or  scattering of the light  induce  radiation pressure forces  which may be either resonant \cite{Minogin:1979} or nonresonant. The nonresonant force acts on the atom or molecule when the frequency of the light wave is far from any resonant frequency of the particle. This force arises due to the Compton scattering of a photon. 
 
V.S. Letokhov showed in \cite{Letokhov:1968} that the neutral atoms can be trapped by means of nonresonant forces.
Namely for standing  plane wave  at intensities    $\gtrsim 10^3-10^4\ \Wt/\cm^2$ the atoms  moving at small angles to the wavefront  can be caught between nodes and antinodes of the standing wave owing to the dipole or striction force 
 \begin{equation}
f=\frac{1}{2} k_{\omega} \vec{\nabla} (\vec{E}^2).
\label{dipoleforce}
\end{equation}
  $\vec{E}$ is the electric field strength, $k_{\omega}=(n_{\omega}-1)/(2\pi N)$ is the atomic polarizability at the frequency $\omega$ of the external light field, $n_{\omega}$ is the   refraction index of an atomic gas, $N$ is the density of particles.
So a configuration of standing waves in three mutually perpendicular spatial directions can serve  the optical trap for atoms. However the number of particles moving at not large angles to the wavefront is small  and the trapping requires significant costs of energy. 
 
When the wave frequency is in  resonance with  an atomic spectral transition, the resonant force of light pressure appears, it may arise due to spontaneous or induced radiation \cite{Letokhov:1977}.  

In the case of spontaneous radiation the   force  arises due to the momentum recoil when the atom absorbs photons from  an incident light   wave.
After the absorption of a light photon the excited atom emits another photon but in a random direction coming back to the ground state.
An average contribution to the recoil force from  the emission processes is zero because spontaneous irradiation can occur in any spatial direction.

The  dipole force  appears in the field of light standing wave during   stimulated  absorption and emission of photons by atoms
along the direction of the incident  wave. 

In 1970 A.\,Ashkin studied the deflection of sodium atomic beam by light pressure forces \cite{Ashkin:1970a} and
obtained the formulas for  scattering   \cite{Ashkin:1970b,Ashkin:1978} and  dipole \cite{Ashkin:1978} forces  acting on the atom in the field of  a monochromatic light wave taking into account the saturation effect.
   V.S.\,Letokhov \cite{Letokhov:1977:2} and  A.\,Ashkin  showed that  the scattering force alone is not sufficient to capture    atoms \cite{Ashkin:1984}, but it may  accelerate  \cite{Ashkin:1970a,Kazantsev:1972,Kazantsev:1974}  or slow down    the atoms \cite{Hansch:1975,Chu:1985}.

Consider  a two-level atom. It absorbs a photon with the momentum $\hslash \vec{k}$ and then reradiates  another photon. 
The Doppler shift $\triangle \nu = p/(m \lambda)$  is less than the  natural linewidth $\gamma =1/ \tau$ and we can average  over a large number of  radiation and absorption processes. So the spontaneously irradiated photons do not contribute to the average scattering force. 

The  resonant force acting on the atom in the optical molasses has the following form 
\begin{equation}
\vec{F}=\frac{\hslash \vec{k}}{2\tau} \frac{I/I_0}{1+I/I_0+\left(   2 \delta \nu / \gamma   \right)^2} \ ,
\end{equation}
where $\delta \nu = \nu - \nu_0 - \vec{k} \vec{v}/(2 \pi)$ is the  frequency detuning of the laser wave from resonance in the   reference frame of the atom, 
 $\nu$ is the frequency  of the  incident light, $\nu_0$ corresponds to the resonant frequency of an atom moving with velocity $\vec{v}$, $k=2\pi/\lambda$ is the wave vector.  
 $I$ is the intensity of the laser field, or energy flow density. $I_0$ is the saturation intensity  
 \begin{equation}
 I_0=\frac{4 \pi^3 \hslash c \gamma}{3 \lambda^3},
 \end{equation}
which corresponds to the irradiance when the  level populations of the ground and excited states are equal.

Consider the case of two waves propagating in the opposite directions with the same frequency and intensity, the intensity of the waves is much less than the saturation intensity $I/I_{0} \ll 1$. The total force is   the sum of the scattering forces acting from the each  wave and for $kv\ll \delta \nu$, $kv\ll \gamma$ has the following form

\begin{equation}
F=4 \hslash k \frac{I}{I_{0}} \frac{kv (2 \delta)/\gamma}{[1+(2 \delta \nu /\gamma)^2]^2}.
\label{eq11}
\end{equation}

When the laser frequency detuning is positive $ \delta \nu > 0$,  atoms
accelerate and heat. For the  negative frequency detuning $\delta \nu <0$ the atoms  slow  down.
The force considered here is equivalent to the viscous friction force $ F = - \alpha v $ with the coefficient

\begin{equation}
\alpha=-4 \hslash k^2 \frac{I}{I_{0}} \frac{ (2\delta \nu)/\gamma}{[1+(2\delta \nu/\gamma)^2]^2}.
\label{eq12}
\end{equation}

This analogy leads  to the new conception of "optical molasses". Due to the absorption and emission  of laser photons an atomic movement looks like the Brownian motion in a region restricted by laser beams \cite{Chu:1985}. The nonzero viscosity slows down the atoms in this region, however the  optical molasses is not a trap and the viscosity force alone is not enough to capture   neutral atoms.

\subsection{Doppler cooling}
\label{subsec:4}

An atom at rest equally absorbs  light from both waves. At non-zero temperature the atom participates
in the thermal motion.
An atom moving with the velocity $v$ sees the  frequency of backward wave shifted to a higher frequency region by  amount of Doppler shift $|\vec{k}\vec{v}|/(2\pi)$. We do not consider the second order Doppler effect  here.

At $\delta \nu<0$ an atom absorbs more photons from the counter-propagating wave than from the collinear one because  the interaction cross section is larger for the backward wave.
Spontaneous emission of  light by an atom occurs in a random direction, and the average recoil force  is zero.
When the wave vector of a photon is  opposite to the atom momentum, the atom losses  energy and slows down \cite{Phillips:1982}. This method of atoms deceleration is called the Doppler cooling \cite{Letokhov:1976b,Wineland:1979,Gordon:1980}.

The cooling process considered here competes with the phenomenon of random walks, i.e. the diffusion of atoms in   optical molasses.
At first the conception of optical molasses was introduced in \cite{Chu:1985} because an atom in the  laser field moves like a particle in a viscous medium.

When the atom emits a photon randomly in an arbitrary direction it receives  the recoil momentum $\hslash k$.
The average atomic momentum $\langle\vec{p}\rangle$ doesn't vary, but its mean square $\langle(\vec{p})^2\rangle$ changes
for the value of  $\hslash^2 k^2$ after an each photon emission. This results to an increase of the average kinetic energy of  the atoms, i.e. heating.
There is a certain temperature limit of the Doppler cooling determined by  the  configuration of   laser beams and magnetic fields in a cooled volume.

D.Wineland and  V.Itano  calculated in  one-dimentional case \cite{Wineland:1979} the Doppler limit when  $|kv|\ll |\delta|,\ |kv|\ll \gamma,\ I/I_0\ll 1$ both for the case of free atoms and atoms  subjected to the external magnetic field.
The cooling rate is determined by the decrease rate of the particle kinetic energy
\begin{equation}
\left(\frac{dE}{dt}\right)_{\cool}=Fv=-\alpha v^2,
\label{eq19}
\end{equation}
where the coefficient $\alpha$ is defined by (\ref{eq12}).

The  heating rate
\begin{equation}
\left(\frac{dE}{dt}\right)_{\heat}=  \frac{\hslash^2 k^2}{M} \frac{\gamma I/I_0}{1+(2\delta \nu/\gamma)^2},
\label{eq21}
\end{equation}
where $M$ is the mass of an atom.
The minimal temperature of the Doppler cooling is determined by the equal   rates of heating and cooling   $(dE/dt)_{\heat}=-(dE/dt)_{\cool}$. Therefore taking into account the equality $m\langle v^2 \rangle/2=k_B T/2$  we get
\begin{equation}
 T=\frac{\hslash \gamma^2} {8 k_B |\delta \nu|} (1+(2\delta \nu/ \gamma)^2),
\label{eq23}
\end{equation}
where $k_B$ is the Boltzman constant. 
The  minimum temperature can be found from \eq{eq23} keeping  the frequency detuning equal to the  half of the level width $ \delta \nu= - \gamma/2 $ during cooling, then the Doppler limit is equal to

\begin{equation}
T_{\Dop}=\hslash \gamma/(2 k_B).
\label{eq23:a}
\end{equation}

One can find the Doppler limit for   three orthogonal laser beams \cite{Phillips:1989}, the cooling rate will be the same as in one-dimensional case however the heating rate will be in three times larger. Since $m\langle v^2 \rangle_{3D}/2=3 k_B T/2$ then we get the same Doppler limit \eq{eq23:a} as in the one-dimensional case.

\subsection{Role of dipole force}
\label{subsec:5}

Consider a neutral atom in an external light wave. Electrons of the atom oscillate under the influence of an  alternating  electric field and irradiate electromagnetic energy. If the intensity of the light field is large enough then   the atom radiates coherently  with the external light field. Such radiation is called stimulated or induced. During radiation dipole or gradient force acts on the atom \eq{dipoleforce}.

At small intensity of laser wave ($I/I_0\ll 1$) the dipole force is much smaller than the scattering force and usually can be neglected. However at very small temperatures, when the Doppler cooling doesn't work anymore, the atoms can be cooled owing to the dipole force, which acts even on the atoms at rest. This is subdoppler cooling and we discuss its mechanisms in the next section.

Strong dipole force acts on the atom in a very intense light wave at $I/I_0\gtrsim 1$.  The fast laser cooling due to dipole force is possible in a standing wave created by two counterpropagating laser waves of high intensity. The atom can absorb  the photon from one laser wave and reradiate it in another laser wave. Such interaction with the both plane waves leads to the appearance  of   dipole force.
  
Jan Dalibard in his PhD dissertation gave a precise expression for the damping coefficient $\alpha$ for an atom moving in a standing wave. This result was obtained with help of perturabative approach \cite{J.Dalibard:1986}. It is valid if $|kv|\ll \gamma$ and has the following form 
\begin{equation}
\alpha=-\frac{\hslash k^2}{(1+4s)^{3/2}} [\frac{4s(2\delta \nu/ \gamma)}{1+(2 \delta \nu / \gamma)^2}(1+2s)
\label{eq14}
\end{equation}
$$
-\frac{2 \delta}{\gamma} [1+6s+6s^2-(1+4s)^{3/2}] ],
$$
where $
s=\frac{I/I_0}{1+(2\delta \nu/\gamma)^2}
$
is the saturation parameter.

To eliminate the effects of simultaneous interaction with both plane waves one can consider two
counterpropagating waves with different  $\sigma^{+}$ and $\sigma^{-}$  polarizations (relative to each other the polarization vectors will rotate in the same direction).
The only transition allowed is  $J=0\rightarrow J=1$. The  damping coefficient has the following form
\begin{equation}
\alpha=-\frac{2 \hslash k^2  s}{1+2s} \frac{2(2\delta \nu/\gamma)}{(2\delta \nu/\gamma)^2+[1+0.5s[(2\delta/\gamma)^2+1]]^2}
\label{eq16}
\end{equation}

\begin{figure}[htb]
\begin{center}
 \includegraphics[scale=0.9, angle=0]{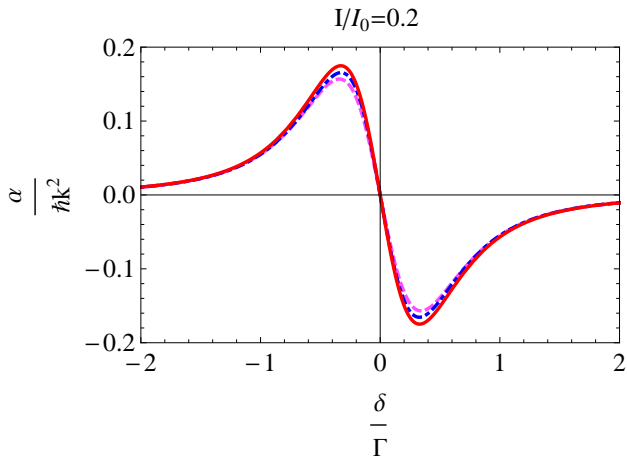}
\caption{Damping coefficient in units $\hslash k^2$ versus  detuning of the laser wave for classical molasses (solid), standing wave (dashed-dotted), and $\sigma^{+} - \sigma^{-}$ (dashed) for $I=0.2I_0$.}
 \label{fig:1}
\end{center}
\end{figure}

\begin{figure}[htb]
\begin{center}
 \includegraphics[scale=0.9, angle=0]{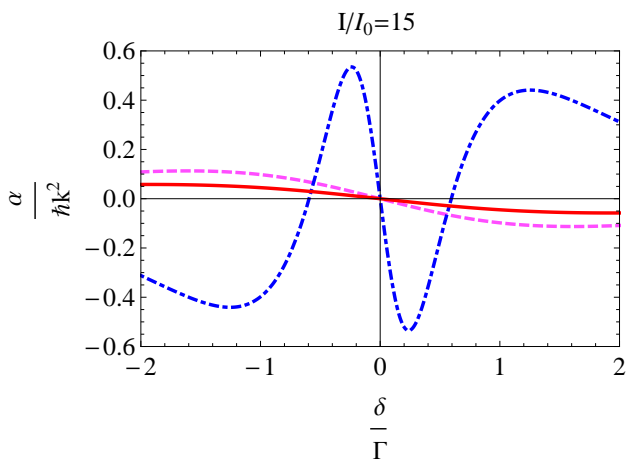}
\caption{The same that on Figure \ref{fig:1}, but for $I=15I_0$.}
 \label{fig:2}
\end{center}
\end{figure}

In Figures \ref{fig:1} and \ref{fig:2} we see the  illustration to this behaviour.
The damping coefficients $\alpha/(\hslash k^2)$ are shown versus the value of detuning.
 
The solid lines correspond to the one-dimensional classical molasses, the dashed-dotted lines represent the damping coefficient for a standing wave, the $\sigma^{+} - \sigma^{-}$ wave configuration  is depicted by the dashed lines.

At the intensity   $I=0.2I_0$ all three cases almost do not differ, because dipole forces are small.
Red detuning $\delta \nu<0$ provides cooling and slowing of atoms, blue detuning $\delta>0$ leads to heating.

At the intensity $I=15I_0$ the field of   standing wave results in  much faster cooling than  classical molasses or the $\sigma^{+} - \sigma^{-}$ wave configuration. In a standing wave at high intensity the atom  subjected  to a very strong damping force either for  blue detuning at $\delta \nu\gtrsim 0.6 \Gamma $ or for  slightly red detuned laser wave. 

Therefore dipole forces  also can lead to the heating of neutral atoms at negative laser wave detuning.
J.Dalibard and C.Cohen-Tannoudji explained this effect   due to a concept of dressed states \cite{Dalibard:1985}.
They suggested to consider an atom interacting with both counterpropagating waves. For comparison in the case of classical molasses at red detuning the atom interacts with counterpropagating running wave, at blue detuning with codirectional laser wave mainly, see previous Section.

\subsection{Conception of dressed states}
\label{subsec:6}

In the approach of "dressed states" the energy levels of combined atom-laser wave system are considered.
This approach is valid if the coupling between atom and laser wave is large, i.e. the Rabi frequency $\omega_1 \gg \gamma$, where  $\gamma$ is the natural linewidth of the excited level and $\omega_1=\vec{d} \cdot \vec{E}/\hslash$, $\vec{d}$ is the atomic electric-dipole moment. This theoretical model is in agreement with experimental results \cite{Aspect:1986}.
 
 Consider a semiclassical approximation, when the size of the atomic wave packet   is small compared with the laser wavelength $\triangle r \ll \lambda$ and $k \triangle v \ll \gamma$.  The frequency detuning  
 $
\delta=\omega_L-\omega_0\ll \omega_L, \omega_0,
$
where $\omega_L$ is the frequency of laser wave, $\omega_0$  is resonance frequency of atomic transition. 
The quantum-mechanical Hamiltonian of atom-laser wave system in dressed state approach has the following form
 
\begin{equation}
H_{DA} (\vec{r})=\hslash (\omega_L - \delta) b^{+} b  + \hslash \omega_L a^{+}_L a_L - 
\label{eq:dressed}
\end{equation}
$$
-[d \cdot \vec{{\cal E}}_L(r) b^{+}a_L +d \cdot \vec{{\cal E}}^*_L(r) b a^+_L ].
$$
The kinetic energy of the atom $\vec{P}^2/2m$   is not considered here. 
In \eq{eq:dressed} the first term corresponds to the internal atomic energy, the second term is the energy of radiation field,
the third one expresses the interaction between atom and laser field.  
  $a^+$ and $a$ are the destruction and the creation operators of a photon,   $b=|g\rangle \langle e|$  and $b^+=|e\rangle \langle g|$ are lowering and raising operators,respectively,  $g$ denotes ground atomic state, $e$ corresponds to the excited state.
  
In \eq{eq:dressed}    $\vec{{\cal E}}_L(\vec{r})$ and $\vec{{\cal E}}^*_L(\vec{r})$ are defined through the positive and the negative frequency components of the electric field  
\begin{equation}
\vec{E}^+(\vec{r})=\vec{{\cal E}}_L(\vec{r}) a_L, \ \ \  \vec{E}^-(\vec{r})=\vec{{\cal E}}^*_L(\vec{r}) a_L^+.
\end{equation}
  
At high intensity the last term in \eq{eq:dressed} has to be taken into account, because it is responsible for the interaction of atomic levels with the electric field of laser wave. 
This interaction is defined by the  matrix element 
\begin{equation}
\frac{2}{\hslash} \langle e,n \vert V   \vert g, n+1   \rangle =-2 \sqrt{n+1} \frac{d {\cal E}}{\hslash} = 
\label{eq38}
\end{equation}
$$=\omega_1(\vec{r}) \exp[i\phi(\vec{r})],$$
where   $\phi(\vec{r})$ is the relative phase between  the vector of dipole moment  $d$
and the vector of external electric field  $\vec{E}$. 
 
\begin{figure}
  \includegraphics[scale=0.65, angle=0]{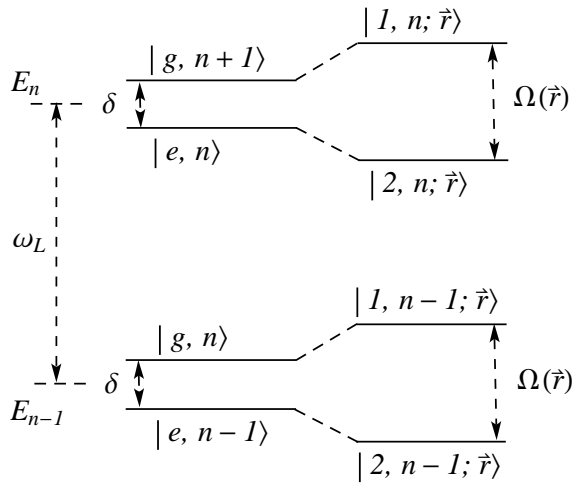}
  \caption{Diagram of energy levels. In the left part of the figure the levels of atom-laser wave system is shown without interaction between laser field and atom. In the right part the dressed states are depicted depending on the atomic position $\vec{r}$ in space.}
  \label{fig4}
\end{figure}

Introduce the concept of generalized Rabi frequency
\begin{equation}
\Omega(\vec{r})=[\omega_1^2(\vec{r})+\delta^2]^{1/2},
\label{eq39}
\end{equation}
then the eigenstates of Hamiltonian of atom-standing wave system have the  form
$$
E_{1n}(\vec{r})=(n+1)\hslash \omega_L - \frac{\hslash \delta}{2}+\frac{\hslash \Omega(\vec{r})}{2},
$$
\begin{equation}
E_{2n}(\vec{r})=(n+1)\hslash \omega_L - \frac{\hslash \delta}{2}-\frac{\hslash \Omega(\vec{r})}{2}.
\label{eq40}
\end{equation}
and shown in  Fig.\ref{fig4} schematically for zero and non-zero Rabi frequency.

For the case of small induced dipole moment (the intensity of laser field is small) the levels picture is presented in left part of  Fig.\ref{fig4}, there are two sets of states corresponding two atomic energy levels and not interacting with laser field of standing wave.
 The following notations are used:  the difference between 
  $E_n$ and $E_{n-1}$ is equal to the energy of one laser photon   $\hslash \omega_L$. Each level has a pare of states: one ground and one excited state.  
For example the level  $E_n$ has one ground state  $|g,n+1\rangle$ with the number of laser photons  $n+1$ 
and one excited state $|e,n\rangle$ with the corresponding number of laser photons  $n$.

For the case of non-zero Rabi frequency we obtain the picture in the right part of  Fig.\ref{fig4}, the states 
 $\vert 1,n; \vec{r} \rangle$ and  $\vert 2,n; \vec{r} \rangle$ at the
 position $\vec{r}$ of the atom. The eigenvectors of hamiltonian 
\eq{eq:dressed}
$$
\vert 1,n; \vec{r} \rangle = \exp[i\phi(\vec{r})/2]\cos\theta(\vec{r}) \vert e,n \rangle
$$
$$
+ \exp[-i\phi(\vec{r})/2] \sin \theta(\vec{r}) \vert g,n+1 \rangle,
$$
$$
\vert 2,n; \vec{r} \rangle = -\exp[i\phi(\vec{r})/2]\sin\theta(\vec{r}) \vert e,n \rangle
$$
\begin{equation}
+ \exp[-i\phi(\vec{r})/2] \cos \theta(\vec{r}) \vert g,n+1 \rangle,
\label{eq41}
\end{equation}
where angle $\theta(\vec{r})$ is defined by the following expressions
\begin{equation}
\cos2\theta(\vec{r})=-\frac{\delta}{\Omega(\vec{r})}, \ \ \sin 2\theta(\vec{r})=\frac{\omega_1}{\Omega(\vec{r})}.
\label{eq42}
\end{equation}

So the dressed states are the superposition of the ground and excited states in the field of electromagnetic wave \eq{eq41}. 
In the absence of laser field the distance between levels is equal to $\hslash \delta$ in energy units. 
Inside the intense laser beam the distance between levels   is equal to  $\hslash \Omega(\vec{r})$ and can be much larger 
than the value of $\hslash \delta$. 
 
 The value of energy gap between the ground and excited states depends on the atom position $\vec{r}$.  
\begin{figure}
  \includegraphics[scale=0.55, angle=0]{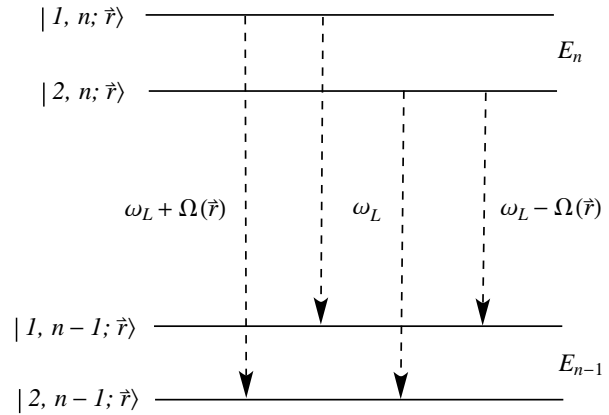}
\caption{Spontaneous transitions between various atomic energy states at some coordinate $\vec{r}$.}
  \label{fig5}
\end{figure}
 Without interaction with intense electromagnetic laser wave there is only one transition in an atom: from   $\vert e,n \rangle$ state to  $\vert g,n \rangle$ state. If the atom interacts with the laser wave, three additional transitions appear. 
 
 The all transitions are shown in Fig.\ref{fig5} by dashed lines. Two transition occur at the 
 same frequency $\omega_L$, other two corresponds to the frequencies  $\omega_L + \Omega(\vec{r})$ and $\omega_L - \Omega(\vec{r})$.  
  
So called triplet Mollow was  obtained in experiment and confirmed the theory \cite{Mollow:1969}.
Results of experiment \cite{Aspect:1986}  also agree with this theory.
 
The expression for gradient force acting on the atom at rest was obtained in  \cite{Dalibard:1985} 
\begin{equation}
f_{\dip}^{\st}=-\nabla \left[ \frac{\hslash \delta}{2}\log\left( 1+\frac{\omega_1^2}{2\delta^2}  \right)  \right]
\label{eq43}
\end{equation}

For atom moving with velocity  $\vec{v}$  the dipole force
\begin{equation}
f_{\dip}(\vec{r},\vec{v})=f_{\dip}^{\st} - \frac{2\hslash \delta}{\Gamma} \left( \frac{\omega^2_1(\vec{r})}{\omega^2_1(\vec{r})+2\delta^2}  \right)^3 (\alpha\cdot v)\alpha,
\label{eq44}
\end{equation}
where  $\alpha=\nabla \omega_1(\vec{r})/\omega_1(\vec{r})$.

For  very cold atoms and small Doppler shift  $kv\ll \Gamma$, J.\,Dalibard derived an expression for the dipole force in a quasi-classical approximation. According to his results at positive laser frequency detuning $\delta>0$ the gas of atoms cools down while the dipole force pushes it out of a high field region, where the scattering force is smaller than the dipole one. Negative frequency detuning $\delta \nu>0$ leads to  heating of atoms, the atoms are accelerated and drawn to the high  field region.

\subsection{Subdoppler cooling}
\label{sec:7}

In 1989   theorists have proposed some models to explain the observed experimental results on subdoppler laser cooling. 
It turned out that the Doppler limit can be overcomed. 
To achieve the temperature much lower than the Doppler limit is possible when the nonadiabatic effects are large for the atom moving in a standing wave.

 It means that the internal state of the atom changes significantly during its motion in a standing wave. 
If the radiative lifetime of the excited state  $\tau_R=1/\gamma$ and  the optical pumping time   $\tau_p=1/\gamma \prime$, where $\gamma \prime$ is the width of the ground state, then $\tau_p \gg \tau_R$ and $\gamma \gg \gamma \prime$. Also at small intensity Rabi frequency $\Omega\ll \gamma$. Therefore the nonadiabatic effects can appear at $kv \sim \gamma \prime$ that is much smaller than the velocity capture range of Doppler cooling $kv \sim \gamma$.

Consider two counterpropagating waves which form a  standing wave. The amplitudes of the waves are equal, their polarizations 
  $\vec{\epsilon}_x$ and $\vec{\epsilon}_y$ a linear and mutually orthogonal.  
  The total field of the standing wave is the vector sum  o $\vec{\cal E}=\vec{\cal E}_1+\vec{\cal E}_2$, where
\begin{equation}
\vec{\cal E}_1={\cal E}_0 \vec{\epsilon}_x (e^{ikz-i\omega_L t}+e^{-ikz+i\omega_L t})
\label{eq48}
\end{equation}
is the electric field of the first wave
\begin{equation}
\vec{\cal E}_2={\cal E}_0 \vec{\epsilon}_y (e^{-ikz-i\omega_L t}+e^{ikz+i\omega_L t})
\label{eq49}
\end{equation}
is the electric field of the second wave.
Their sum has the following form
\begin{equation}
\vec{\cal E}=2 {\cal E}_0 \cos(kz)(\vec{\epsilon}_x+\vec{\epsilon}_y)\cos(\omega_L t) +
\label{eq50}
\end{equation}
$$
+2 {\cal E}_0 \sin(kz)(\vec{\epsilon}_x-\vec{\epsilon}_y)\sin(\omega_L t)
$$
One can see from (\ref{eq50})  that the polarization  
  $\vec{\epsilon}_1=\vec{\epsilon}_x+\vec{\epsilon}_y$ is linear at $z=0$, circular  $\sigma_-$ at  $z=\lambda/8$. At $z=\lambda/4$ the polarization is again linear  $\vec{\epsilon}_2=\vec{\epsilon}_x-\vec{\epsilon}_y$, at  $z=3\lambda/8$ - circular $\sigma_{+}$ and so on.

\begin{figure}[htb]
\begin{center}
 \includegraphics[scale=0.715, angle=0]{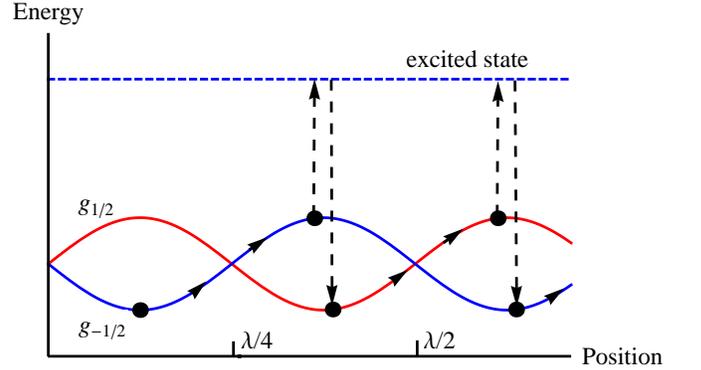}
\caption{Laser cooling by the method of  polarization gradient. The energies of the ground states $g_{1/2}$ and $g_{-1/2}$ in the standing wave depending on the atom position, $\lambda$ is the wavelength.}
 \label{fig:siscool}
\end{center}
\end{figure}

 In Fig.\ref{fig:siscool} the energy of an atom is shown. The atom moves in a low intensity standing wave. 
The  levels $g_{-1/2}$ and $g_{+1/2}$ are ground states with the magnetic quantum numbers $m_J=-1/2$ and $+1/2$ respectively.
The energy of each level changes periodically in space due to its interaction with the laser wave (dressed states). Shift of the atom energy, which appear as a result of this interaction, is called light shift $\triangle E_0$. Various sublevels have different light shift at a given point of space, because the energy interaction between a photon and electron in a given quantum state depends on the  magnetic moment of the photon (or the wave polarization).

\begin{figure}[htb]
\begin{center}
 \includegraphics[scale=0.62, angle=0]{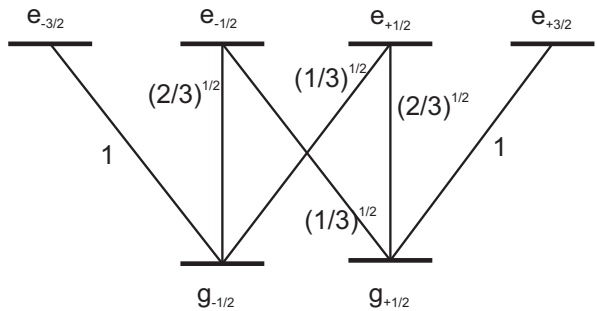}
\caption{Clebsh-Gordan coefficients for the transitions between various sublevels of ground $g$ and excited states for the case when the total angular momentum of the ground level is   $j_g=1/2$ and  the angular momentum of the excited level is  $j_e=3/2$.}
 \label{fig:Clebsch}
\end{center}
\end{figure}

 The probability of atom transition between ground $g$ and excited $e$ states is proportional to the squared 
 Clebsh-Gordan coefficient for the corresponding levels, these coefficients are shown in Fig.\ref{fig:Clebsch} for various Zeeman  sublevels.

The atoms are translated from one ground state  sublevel  to  one of sublevels of an excited state during the
optical pumping time $\tau_{\opt}$. The probability of photon absorption is the largest at  maximal value of energy.
 At  properly selected time of  optical pumping
$\tau_{\opt} \sim \lambda/(4 v)$ and $ \tau_{\rad} \ll \tau_{\opt}$  atoms cool down in the standing wave,
because the atom climbs the hills of potential energy more often than goes down losing the energy.

Obviously the minimum temperature in this method is determined  by changing  potential energy of an atom moving in the standing wave. The maximum change called  light shift is

\begin{equation}
\triangle E_0 \simeq  k_B T_{\mmin}.
\label{eq53}
\end{equation}

For small recoil energy and light shift much smaller than the energy width of the excited state $\hslash \Gamma$ it is possible to get the temperatures of the micro Kelvin range \cite{Phillips:1989,Lett:1988}.
So at decreasing intensity of light wave the temperature of atoms decreases also  until it reaches some limit determined by  single-photon recoil energy $E_{\rec}$ (\ref{eq25})
\begin{equation}
 T_{\rec}=\frac{\hslash^2 k^2}{2M k_B},
\label{eq53}
\end{equation}
where $M$ is the atom mass.
The limitation arises, because the last spontaneously emitted photon has a finite momentum $ \hslash k $ .
The subdoppler cooling  works effectively due to the dipole force in a near-zero velocity range  \cite{Dalibard:1989}. For sufficiently large velocities  it  becomes ineffective, and  the Doppler cooling comes into play.
The both methods may be used for  laser cooling of antihydrogen.

 To get the  temperatures below the Doppler limit is possible in a standing wave, but neither with $\sigma^{+} - \sigma^{-}$ configuration nor with classical molasses.
 
According to the of subdoppler cooling  \cite{Dalibard:1989} the temperature of atoms   depends on  detuning in the following way

\begin{equation}
 T = C \frac{\hslash \omega_1^2}{|\delta \nu| k_B},
\label{eq45}
\end{equation}

where  $C = 1/8 $, $\omega_1$ is  Rabi frequency,  $\delta=\omega_L-\omega_0$ is  frequency detuning. This theoretical behaviour agrees with the  experiment \cite{Solomon:1990}.

It is known that from the experiments with laser waves of parallel polarizations the coefficient equals $C_{\|}=0.45$ \cite{Solomon:1990}, and for  perpendicular laser polarizations  $C_{\bot}=0.35$ \cite{Solomon:1990} obtained for six laser beams.
The theory of cooling  predicts  $C_{\bot}=0.13$ for two counterpropagating waves \cite{Dalibard:1989}.
The experiments showed that the  minimum achievable temperature is also proportional to the intensity of a laser wave \cite{Solomon:1990}.

In a magnetic trap the rate of laser cooling  depends on the magnetic field. This dependence is easily explained by  Zeeman shift of atomic levels. To get effective cooling in the magnetic field the Zeeman splitting and shifts of the  levels in a magnetic field should be taken into  account when we choose an optimal frequency detuning \cite{Phillips:1989}. Neglecting the effects of magnetic field  causes to slow  cooling and spurious optical pumping into untrapped states.

\section{Review of experiments}
\label{sec:8}

 \subsection{Precision  measurements. Spectroscopy.}
 \label{subsec:9}
 
 A great work has been already done and 
the incredible precision was achieved in measurements of atomic hydrogen spectrum.
With an improved detection method the frequency of $1S-2S$ transition was obtained with a  relative accuracy $4.5\times 10^{-15}$ and 
it  equals to $2466061413187018(11)\ \Hz$ \cite{Matveev:2013}. 
 The $1S$ ground-state Lamb shift in hydrogen atom  was measured in 1992 with the relative accuracy  equal to  $1.3\times 10^5$ and its value is $8172.82(11)\ \MGc$ \cite{Weitz:1992}, 
the experiment was performed at room temperature by Doppler-Free Two-Photon spectroscopy.
The frequency of  $GS-HFS$ splitting of $1S$ level in hydrogen atom is known with $6.3 \times 10^{-13}$ and equals to $1420405751.7667(9)\ \Hz$  \cite{Essen:1971}, $2S$ hyperfine interval has a value  $177556834.3(6.7)\ \Hz$ \cite{Kolachevsky:2009b}.

There is a hypothesis about a drift of coupling constant $\alpha$ in time \cite{Calmet:2002} based on astronomical observations of spectral lines \cite{Murphy:2003}.
 Annual measurements of atomic transition frequencies make possible to determine a change of the coupling constant $\alpha$
 \cite{Kolachevsky:2009}. If this change is not zero then  in accordance  with the  Grand Unification Theory   other physical  coupling constants are also changing in time.
However such changes of fundamental constants are forbidden in theories with a metric including general theory of relativity, because it violates the Einstein's equivalence principle.  

Previous data on the $1S-2S$ transition frequency \cite{Huber:1999,Fischer:2004} allow to set a limit on the $\alpha$ coupling constant drift.
The drift equals to $(3.2\pm 6.3)\times 10^{-15}$ per year, i.e. zero in the range of errors.
The precision spectral measurements of hydrogen and antihydrogen are important because they could  set a limit  on an applicability of modern physics laws. 
The results were obtained by use of the optical frequency comb formed by a femtosecond laser. Optical combs  lead  to a revolution in a high precision measurements, for detailed study we recommend a  review \cite{Hansch:2006}.

The one of the  primary goals  of $CERN$ collaborations is a  high precision laser spectroscopy of antihydrogen atom.
Today the accuracy of such experiments is low, experimentalists have to work  with  small amount of antiatoms.  The problems are often associated  with the arrangement of an experimental setup. 
For example  $ALPHA$ collaboration has measured the ground-state hyperfine splitting $(GS-HFS)$ of antihydrogen atom  \cite{Amole:2012:HFS}. To avoid a loss of antiatoms due to annihilation the experiment was done in the magnetic trap with the depth $0.5\ \K$. The experimental measurements were performed with the relative accuracy $3\times 10^{-3}$. 
There are various uncertainty sources such as the low statistics, magnetic field of the trap, etc.

 Hyperfine splitting of energy levels is mainly a magnetic effect and occurs due to the interaction of the magnetic moments of positron and antiproton. 
To achieve a high precision in  study of the ground-state hyperfine splitting  the magnetic field should be absent.  
It was suggested to use a "cusp" configuration of magnetic fields for the synthesis of  atomic antihydrogen and further spectroscopic measurements \cite{Kuroda:2014}. The scheme   is described in \cite{Mohri:2003,Enomoto:2010}. 
 Coalescence of positrons and antiprotons occurs in-flight and the spin-polarized beam of antihydrogen atoms goes out of the cusp trap. 
  Using this method $ASACUSA$ collaboration obtains in $5-50$ times larger antiatoms than in experiments of $ATRAP$ and $ALPHA$ during one working cycle of antiproton decelerator \cite{Kuroda:2014}. In the radio-frequency quadrupole decelerator $(RFQD)$   antiprotons are slowed down  to $115\ \Kev$   before a movement to the cusp trap for the antihydrogen synthesis.
To increase the production of antitoms the rf-drive is applied to antiprotons.  It turns possible to get $25$ antiatoms per hour with the principal quantum number  $n\lesssim 29$ for  number of antiprotons $5\times 10^7$ coming from Antiproton Decelerator. ASACUSA aims to measure the $GS-HFS$ splitting of antihydrogen using a beam of antiatoms from the cusp trap \cite{ASACUSA:HFS}.

Laser cooling could help to improve an  accuracy of some experiments on the spectrum of hydrogen and antihydrogen.
Modern  Doppler-Free spectroscopy avoids errors due to the first-order Doppler broadening. But there is the second order Doppler effect, which shifts a spectral line, changes its shape and  degrades an  accuracy of   experiments. 
  If measurements are performed in a magnetic trap, the external magnetic field leads to the   magnetic broadening of the spectral lines.  The colder   atoms localize in the region of the trap with the weaker   field,  thus  the    accuracy of the experiment is higher.

  As a powerful source of Lyman radiation is not still created, the experiment with a beam of antimatter seems more promising, since the measurements can be performed in the absence of the magnetic field. Powerful laser radiation could also be used for collimation of an experimental beam, see Section 8.

 At  high temperatures  the energy exchange can occur between internal and external degrees of freedom in an atom owing to collisions, which also leads to a shift of atomic energy levels. At thermal energies the level broadening  $\triangle \nu \sim  n \sqrt{T}$, where $n$ is the concentration of atoms,   $T$  is the temperature. For   number of antihydrogen atoms which we have in the experiment this broadening is absent.

The  number of  antiatoms has to be increased. Typically, the spectroscopic measurements are performed with millions of neutral atoms within fractions of  a second, or about a second. If the number of antiatoms would be  the same as in  experiments with usual atoms, then  the spectra have been able to measure during the dispersion with a high precision. We do not have so many antiatoms. However, scientists have learned to keep antiahydrogen in the magnetic trap of superconducting type $\sim 1000\ \cc$   at the number of antiatoms of a few hundred. 
   It would be desirable to increase by two orders of magnitude either the retention time or the number of antiatoms.  $ELENA$ experimental facility   would lead to an increase of  antihydrogen yield.

 \subsection{Experimental check of Einstein's equivalence principle} 
 \label{subsec:10}

Another research area in  antihydrogen physics  is the check of Einstein  Equivalence Principle,  the basis of theory of general relativity. The observed acceleration of Universe expansion causes many fundamental questions as our understansing of the gravitation is incomplete.  

 There are many compelling arguments which confirm  the equality of gravitational and inertial mass of matter and antimatter, but they are indirect and based on theoretical assumptions such as CPT theory and Lorentz invariance.
 
$ATRAP$  collaboration  found the ratio  of gravitational  mass to inertial mass $F<200$ at   $2\sigma$ level \cite{Gabrielse:ATRAP:2012}, the measurements were performed in the magnetic trap. The next experiment of $ALPHA$ collaboration showed that   $-65 \leq F \leq 110$ at a statistical level of significance equal to $95\%$ \cite{Charman:2013}. It is also affirmed in this work that cooling the antiatoms to temperatures $\leq 30\, \mK$ and increasing the shut-down time of the magnetic field in the trap   makes possible to measure the ratio $F$ with $\pm 1$ accuracy.
 
The  frequency  analysis of the non-linear antihydrogen dynamics in  magnitostatic trap has been done in \cite{Zhmoginov:2013} using numerical methods. 
It was shown that  stochasticity is insufficient and ergodicity is not satisfied for the antiatom motion in the magnetic trap because radial and axial motions of antiparticles are uncoupled. The accuracy of $F$ determination strongly depends on the shut-down profile of the radial field component. 
The requirement of ergodicity is essential assumption for the gravitational experiments carried out by $ALPHA$ \cite{Charman:2013} and $ATRAP$ collaborations \cite{Gabrielse:ATRAP:2012}.

Collaboration  $AE\bar{g}IS$ proposed to measure the gravitational acceleration of antihydrogen using the shift of interference fringes created by antiatom beam \cite{aegis1,aegis2,aegis3}.  
Atomic  antihydrogen is synthesized via  resonant charge-exchange collisions of Rydberg $Ps^*$ with antiprotons at temperature $100\ \mK$, the number of $10^5$ antiatoms is needed to obtain $\bar{g}$ with  $1\%$ accuracy.

Collaboration $GBAR$ also produces antihydrogen atoms via a resonant charge-exchange collisions but their in-flight measurement of $g$-value demands much lower atomic velocities of the order of $\m/\cc$ or temperatures $\sim 20\, \mu \K$. 
Antihydrogen ions $\bar{H}^+$  are produced  through the reactions 
\begin{equation}
\bar{p}+Ps^* \rightarrow \bar{H} + e^-,
\end{equation}
\vspace{-0.6cm}
\begin{equation}
\bar{H}+Ps^* \rightarrow \bar{H}^+ + e^-.
\end{equation}

Then
$\bar{H}^+$  are sympathetically cooled by beryllium ions to $\mu \K$ temperature range. The second positron is photodetached and the antiatoms fall down on the detecting plate, where the  annihilation takes place.
This production scheme is complicated so the $\bar{H}$ yield  is low. Experimentalists  aim to obtain   antihydrogen atoms initially  at temperatures $\sim \mu \K$.

$GBAR$ also  plans to measure the  gravitational levels of antihydrogen like  for the   neutron \cite{Nesvizhevsky}.  
The advantage  is that   antihydrogen atoms are detected much better than neutrons. The spectroscopy of gravitational levels could improve the accuracy of $\bar{g}$ measurement \cite{Voronin:2013}.

Check of electrical neutrality is very important for the proper interpretation of   gravitational experiments.
Surrounding electric fields   prevent such experiments    with charged particles. 
  The gravitational forces exceed the electrical ones when the electrical charge of the $\bar{H}$ is less than $10^{-7}e$. Results of $ALPHA$ collaboration  show that this condition is satisfied  \cite{Amole:2014}.

$ALPHA$ collaboration set a bound on the electrical charge of antihydrogen   \cite{Amole:2014}. It equals to $Q=(-1.3 \pm 1.1 \pm 0.4)\times 10^{-8}$ with a statistical confidence level of $1\sigma$, the first error is large and caused by small statistics,
 the second one is the consequence of systematic effects.   Modern experiments show that ordinary matter is neutral with  the accuracy  equal to  $(-0.2\pm 1.1)\times10^{-21}e$ \cite{Bressi:2011} and the same has to be true   for antimatter.
 
\subsection{Bose-Einstein Condensation} 
\label{subsec:11}
 
 It was shown in 1976   with many body calculations \cite{Stwalley:1876} that spin-polarized atomic hydrogen retains its gaseous properties at all temperatures.
%!!!!!!!!!!!!!!!!!!!!!!!!!!!!!!!!!!!!!!!!!!!!!!!!!!!!!!!!!!!!!!!!!!!!!!!!!!!!!!!!!!!!%  
%!!!!!!!!!!!!!!!!!!!!   указать во сколько раз меньше, примеры привести  !!!!!!!!!!!!!!!!!!!% 
%!!!!!!!!!!!!!!!!!!!!!!!!!!!!!!!!!!!!!!!!!!!!!!!!!!!!!!!!!!!!!!!!!!!!!!!!!!!!!!!!!!!!% 
 In 1980    at Amsterdam the spin-polarized $H$ was obtained  in experiment by  W. Silvera and J. Warlaven \cite{Silvera:1980}.
Bose-Einstein condensation of spin-polarized atomic hydrogen  is of great interest because the interaction between atoms is very weak, so the measurements of the BEC properties can be performed with a high precision.
  Using atomic hydrogen makes possible to check  the theories  of  condensates  and the theory of ultra-cold collisions, explore  spin-waves, collective phenomena,  correlations, quantum phase transitions,   etc. 
 
Upon reaching the temperatures of the order of microKelvin the experimentalists also have an access to quantum physics with atoms. 
  The ground state of alkali-metal atoms is the solid. It limits the lifetime of such gases  and the accuracy of   experiments.
Hence the use of atomic hydrogen for some experiments may be more advantageous. 

 The  transition temperature of hydrogen to $BEC$ phase occurs at $~50\ \mu\K$. Alkali-metal atoms has a lower critical temperature of condensation, for  example $Li$ condenses at $0.3\ \mu \K$, $Na$ at $2\ \mu\K$. Laser cooling methods allow to precool such atomic species to $\mK$ temperature range.  Laser cooling till $\mu\K$ temperatures can be cariied out only for alkali-metal atoms, but not for atomic hydrogen mainly due to the absence of powefull Lyman-$\alpha$ laser sources.
 
 Many unsuccessful attempts have been undertaken to obtain  $BEC$ phase of hydrogen in laboratory    because  spin-polarized atomic hydrogen recombines into a molecular state owing to the adsorption on the walls and    three-body recombination.
 In 1986 at MIT  hydrogen gas was cooled till $30\ \mu\K$ and density equal to $2 \times 10^{14}$ cm$^{-3}$ due to    evaporative cooling and magnetic compression in a static magnetic trap \cite{Hess:1986}.
 
Bose-Einstein condensate of atomic $H$ was obtained  in 1998  in the magnetic trap and detected using laser spectroscopy \cite{Killian:1998,Fried:1998,Bongs:1999}. The temperature $50\ \mu \K$ and density $1.8 \times 10^{14}$ cm$^{-3}$ were achieved due to the evaporating cooling and  rf  evaporation.  The diameter of the condensate was equal to $15\ \mu \m$ and its length was $5\ \mm$.   The peak condensate density was $4.8\times 10^{15}$ cm$^{-3}$, which correspons to $10^9$ atoms.

 Magnetic trap played a key role in these experiments \cite{Pritchard:1983}. 
Hydrogen atom has a  non-zero magnetic moment and may exist in various spin states. Atoms in the "low-field seeking" state    move to a spatial region   with the smallest field value, i.e. to the center of the trap. Hydrogen is thus isolated from the chamber walls.

$BEC$ phase of antimatter is also extremely interesting for researches. 
 At temperatures close to zero   spin-polarized   antihydrogen would present  an ideal system for   check of quantum mechanics laws. The interatomic interactions can be calculated to  a  high precision.
Unfortunately such low temperatures are far beyond the technical capabilities.
 The antimatter samples cannot    be made   sufficiently  dense at current experimental facilities to apply evaporative cooling.  
  The number of antihydrogen atoms  obtained in a laboratory  amounts to hundreds  today.
 Therefore obtaining of $\bar{H}$ Bose-Einstein condensate   is impossible without laser cooling.  
 Indeed,  unlike conventional hydrogen the antihydrogen atoms cannot be precooled by collisions with  walls of a vacuum chamber,  because it would lead to   annihilation.
 
 Also worth noting   that the atoms in this phase are easy to control, since the scattering length depends on the external magnetic field.

\subsection{Experiments on cooling of hydrogen}
\label{sec:12}

The first and the only experiment on optical cooling of  atomic hydrogen was carried out in 1993 \cite{Setija:1993}.
Hydrogen atoms were  precooled by cryogenic technology to the temperatures  $\sim 80\ \mK$ and loaded into the magnetic trap with the well depth  $\sim 0.82\ \K$.
The   radius and the length of the trap  were $6\ \mm$ and $12\ \cm$ correspondingly. 

Density of  spin-polarized atomic hydrogen at the center of the trap was relatively high $n\gtrsim 10^{11}\ \cm^{-3}$.
The magnetic field configuration in the trap was described in  \cite{Setija:1993,Roijen:1993,Hijimans:1989,Luiten:1993} and its absolute value was determined by
$|B(\vec{r})|=(B^2_{\bot}+B^2_z)^{1/2}$, where $B_{\bot}\approx \alpha \rho$ is the component of the magnetic field  perpendicular to the trap axis, $r$ is the distance from the symmetry axis,  $B_z=B_0+\beta z^2$
is the field component along the axis of the trap, $z$ is the coordinate along this axis, $\alpha=2.2\ \T/\cm$, $\beta=0.023\ \T/\cm^2 $ and   $B_0\simeq 0.1\ \Tl$ is the minimum field value  in the trap.

J. Warlaven and his colleges utilized the Lyman-$\alpha$ transition $1S_{1/2} - 2P_{3/2}$ for laser cooling.
The total power of the laser  source was  $160\ \nWt$. The power in a region of atoms localization was equal to  $2.5\ \nWt$.
Atoms were irradiated by circularly polarized light, its intensity was  100 times smaller than the saturation intensity.
The source generated  $2 \cdot 10^9$ photons per pulse, the duration of the pulse was  $10\, \ncc$, the repetition rate was  $50\, \Gc$. The bandwidth of  laser radiation did not exceed  $100\, \MGc$.

Doppler cooling makes it possible to cool atoms from   $80\ \mK$ to $11(2)\, \mK$ temperature during 15 minutes. Cooling proceeds slowly compared to the inter atomic collisions rate which is equal to  $\tau_c^{-1} \sim 0.1 - 10\ \cc^{-1}$
but much faster than the decay rate due to spin flips in the interatomic collisions $\tau_d^{-1} \sim 10^{-5} - 10^{-3}\ \cc^{-1}$. During  cooling  the density of  atoms increased  16 times, however $99.9994\ \%$ of particles leave the trap due to the spin flips.
Thereafter the Doppler cooling was performed again at   lower density.
The minimum temperature was equal to $8\, \mK$, that is slightly above the  Doppler limit for the hydrogen atom.
Subsequent evaporative cooling makes it possible to get even lower temperatures  $T=3(1)\ \mK$ \cite{Hess:1986,Setija:1993}.

Let's estimate the time of laser cooling in this experiment from  common principles of  Doppler cooling theory.
The magnetic field in the trap is not uniform, so the density of  atoms distributes according to the Boltzmann law and its potential energy is the function of  temperature and an atom position in the magnetic trap is

\begin{equation}
n(\vec{r})=n_0 \exp [-\frac{\mu_B(B(\vec{r})-B_0)}{k_B T}],
\label{eq277}
\end{equation}

\begin{figure}[htb]
\begin{center}
 \includegraphics[scale=0.8, angle=0]{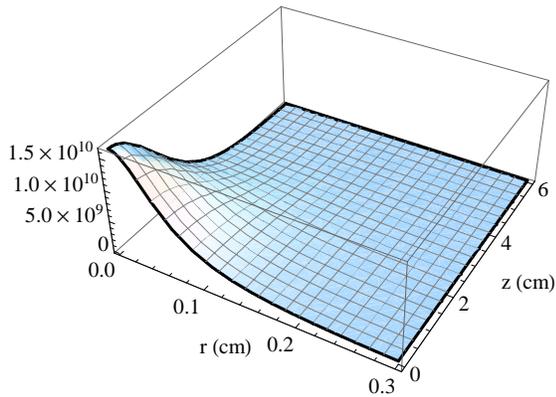}
\caption{Density distribution  of hydrogen atoms in the magnetic trap at $T=80\ \mK$, density at the trap center equal to $10^{11}$ cm$^{-3}$, $r$ is the radial distance from the trap axis, $z$ is the coordinate along the trap axis. }
 \label{fig:density}
\end{center}
\end{figure}

where $n_0$ is the density of atoms at the magnetic trap center, $\mu_B$ is the Bohr magneton, $B(\vec{r})$ is the magnetic field in the point (z,$\rho$).
In Figure \ref{fig:density} we show the density distribution of hydrogen atoms at temperature $T=80\ \mK$ in the magnetic trap.
The total number of particles amounts to  $\sim 3 \times 10^9$ in the trap.

For the most efficient laser cooling the detuning $\delta$ has to be equal $-(\gamma_{sp}+\gamma_{las})/2\approx -6.27 \times 10^8\ \Hz$, where $\gamma_{sp}=6.26\times10^8\ \Hz$ is the spontaneous decay rate of the $2P$ state, $\gamma_{sp}=2\pi \times10^8\ \Hz$ is the laser linewidth. 
 
Using the  Doppler force acting on hydrogen  atom   roughly evaluate the cooling time $80\ \mK$ till $11\ \mK$.
Energy of the atom change with the rate
\begin{equation}
\frac{dU}{dt}=\frac{t_p}{t_R}\frac{I}{ I_0}\frac{\hslash k^2/M_H}{1+ (2\delta/\gamma)^2} \left(\frac{24k_B T \delta}{\gamma(1+(2\delta/\gamma)^2)}+\hslash \gamma\right)
\label{eq277}
\end{equation}
where $k=2\pi/\lambda$, $\lambda=121.56\times 10^{-9}\ \m$ is the laser wavelength, $M_H=1.67\times 10^{-27}\ \kg$  is the mass of atomic  hydrogen, $k_B=1.38\times 10^{-23}\ \Dg/\K$ is the Boltzmann constant.  
 The diameter of   laser beam is equal to $2\ \mm$. We found the cooling time equal to $\approx 10$ min for the intensity in $100$ times smaller than saturation intensity   $I/I_0 \approx 0.01$ \cite{Setija:1993},  while the experimental value amounts $15$ min. The shifts of levels due to the external magnetic field have not been taken into account. The duration of the pulse $t_p=10\ \ns$ and its repetition rate $1/t_R=50\ \Hz$.
  At the initial temperature $80\ \mK$ only $34.07\ \% $ of  all the particles are inside the laser beam. As the temperature reduces the hydrogen atoms localise near the center of the trap and  at $11\ \mK$ the laser beams covers  $99.98\ \%$ of the  particles remaining in the trap.

 For  continuous wave laser source of the power  $20\ \nWt$ the cooling time  would be equal $~110$ min so the pulsed lasers are still the best choice for  cooling of  antihydrogen atoms. 
 CW lasers are necessary for creation of optical molasses or beam-stopping experiments because in the pulsed irradiation with   low pulse duration the atoms (antiatoms) escape simply from the volume in the time between two pulses.
 
 \section{Problems of laser cooling of (anti)hydrogen}
\label{sec:13}

\subsection{Large Doppler limit and recoil energy}
\label{subsec:14}

We estimate some parameters specific for the Lyman transition  $1S_{1/2} - 2P_{3/2}$ in an antihydrogen atom.
Using the value of the transition width, the values of Boltzmann and Plank constants we get from (\ref{eq23:a}) the
Doppler limit $T_{\Dop}=2.4\, \mK$ for the antihydrogen atom. The average velocity of the antiatoms at this temperature   $v=\sqrt{3k_B T/m_{H}}\approx 7.71\, \m/\cc$.  In the experiment due to spatially inhomogeneous shift of Zeeman sublevels the minimal temperature  is usually higher.
The Zeeman shift  of the energy of  excited states 
\begin{equation}
\triangle E= \pm g_J \mu_B B,
\end{equation}
where $m_J\neq 0$ is the magnetic quantum number, $g_J$ is the Lande factor of the excited state, $\mu_B$ is the Bohr magneton.

For  effective Doppler cooling the recoil energy $E_{\rec}$ should be small in comparison with the level width in  energy units $\hslash \gamma$

\begin{equation}
E_{\rec}= \frac{\hslash^2 k^2}{2m_H}\ll \hslash \gamma,
\label{eq25}
\end{equation}

 where $k=2\pi/\lambda$ is the wave vector.
The inequality holds at $k \triangle v \ll \gamma$ in the semiclassical approximation implying $\triangle r \ll \lambda$ where   $\triangle r$ is an atom size.
Taking into account the Heisenberg ratio it's easy to verify that the expression  (\ref{eq25}) holds.
We substitute the numerical values for the Lyman-$\alpha$ transition in (\ref{eq25}) and find
$ E_{\rec} \approx  8.86\cdot 10^{-27}\, \Dg $,  $\hslash \gamma \approx 0.67 \cdot 10^{-25} \Dg $, where $\hslash=1.05\times 10^{-34}\ \Dg\cdot \cc$ so the ratio $E_{\rec}/(\hslash \gamma) \sim 0.13$.

The recoil energy of  antihydrogen atom for Lyman-$\alpha$wavelength  corresponds to the temperature $T_{\rec}=E_{\rec}/k_B\simeq 0.64\, \mK$, the corresponding atomic velocity is equal to $3.98\, \m/\cc$.
The velocity mean change  after  scattering of a single photon  $\triangle v=2 \pi \hslash \nu/(m_H c)\approx 3.26\, \m/\cc$,
 where $m_H$ is the mass of an antihydrogen atom. 
Large recoil energy also slows down the cooling rate and this is not good especially for antiatoms. Their confinement time  is limited because of collisions between antiatoms and   residual  hydrogen $H_2$ and helium $He$ in the magnetic trap. 

All the spectrum transitions in $\bar{H} (H)$ from  the ground state lie in the deep ultraviolet, and have even lower wavelengths and larger recoil energies than for the $1S - 2P$ transition.
Note that for $1S\rightarrow 2S$  two photon cooling scheme the recoil temperature equals  $1.2\ \mK$ \cite{Zehnle:2001}.
Obviously the use of different spectral transitions don't help to overcome the problem.
Nevertheless to cool  antihydrogen atoms below the recoil energy is possible theoretically due to  coherent population trapping, which has already  been  demonstrated for helium atoms  \cite{Aspect:1988,Doyle:1991}. However it needs  high laser power close to saturation and very narrow laser linewidth \cite{Aspect:1988}.

\subsection{Limited optical access}
\label{subsec:15}

In  experiments   atomic collisions with the walls of  vacuum chamber, covered by  liquid helium, assisted significantly in the  achievement of low temperatures $\sim 4\ \K$.  Obviously for further cooling the atomic sample should be isolated from the experimental setup. 
 Collisions with   walls of vacuum chamber    promote the establishment of thermal equilibrium, 
 but lead to  recombination of atoms with the surface. 
 So   the magnetic trap is utilized  to avoid a heating due to collisions with the experimental setup and losses because of recombination. It's  also necessary for the obtaining of spin-polarized atoms and study of its quantum properties. 
But the   design of the magnetic traps used for $H$ and $\bar{H}$ permits the use of only two laser  beams along one spatial axis.
 
Exploring antihydrogen
$ALPHA $ and $ATRAP$ collaborations  use superconducting octupole magnetic trap  with the depth  $ \sim 0.5\ \K $.
It isolates   the antimatter from the walls of a vacuum chamber to prevent  annihilation. The design of the magnetic trap for atomic antihydrogen is more complicated than for hydrogen but both belongs to the same Ioffe type.
Octupole magnetic traps also assume laser cooling of (anti)atoms along the one spatial direction.

From Section \ref{subsec:4} it follows  that  cooling in the longitudinal direction leads to a heating in the perpendicular ones, because  atoms  absorb  laser photons  along  the  beam  axis,  while the spontaneous emissions occur in   random directions.
This presents a problem in the case of low density of (anti)atoms in the trap, because its rare collisions   do not provide   the necessary energy exchange between various spatial degrees of freedom.
  
Density of antihydrogen atoms  obtained at $CERN$ is very small.  
 There is no simple mechanism which can help in achievement of thermal equilibrium because  antiatoms should be  isolated from the trap walls. 
 The possible solution of this problem was proposed in \cite{Donnan:2012}.
It was  suggested to use   nonharmonic magnetic fields to improve a connection between various degrees of freedom.
 The use of a pulsed  source of Lyman-$\alpha$ irradiation with the power of $0.1\, \mDg$,  pulse duration  $10\, \ncc$,  repetition rate  $10\, \Gc$ and   diameter  $10\, \mm$ could give the lowest temperature    $\sim 20\, \mK$.
 
\subsection{Absence of powefull Lyman source}
\label{subsec:16}

The lifetime of antihydrogen in a magnetic trap is not long enough because the vacuum in the magnetic trap is not ideal. So antihydrogen annihilates with the background gases which are present at the experimental setup.
At high density $\sim 10^{13}-10^{14}\ \cm^{-3}$  the processes of spin flips or Maorana transitions come into play because of  the mutual interatomic  collisions   \cite{Lagendijk:1986,Stoof:1988} which are  relevant only for hydrogen experiment for now. 

Spin flips also occurs owing to light  scattering by atom and photoionization processes. 
Such events are  undesirable and  lead to a loss of (anti)particles from the magnetic trap.
Because the lifetime of antihydrogen is limited to several hundreds of seconds  the powefull laser sources are necessary for the rapid cooling of  antiatoms. 

The Lyman-$\alpha$ transition $1S_{1/2}-2P_{3/2}$
in the antihydrogen atom is the best suited for the laser
cooling, because $2P$ state has a shot lifetime equal to $\tau \approx 1.6\, \ncc$. The photoionization probability of excited state $2P_{3/2}$ is small. At saturation regime the cycle absorption-emission is about $5\, \ncc$ due to a very small lifetime of $2P_{3/2}$ state. Spin flips  or Maorana transitions due to the light scattering also occur less frequently on this transition.

However laser sources at  the wavelength $\lambda=121.6\ \nm$ do not have enough power for  fast and efficient cooling of (anti)hydrogen.
The production of continuous
irradiation is very complex and accessible power of such laser
sources is very small today \cite{Scheid:2009,Kolbe:2011}. 
The pulsed  sources of Lyman-$\alpha$ irradiation has a slightly higher power than continuous one.

 Let's estimate  saturation intensity  for $1S_{1/2} - 2P_{3/2}$ transition in antihydrogen atom
 
 \begin{equation}
 I_0=\frac{4 \pi^3 \hslash c \gamma}{3 \lambda^3}.
\label{eq18}
\end{equation}

Its wavelength $\lambda=121.6\, \nm$, the laser frequency  $\nu=c/\lambda= 2.47  \cdot 10^{15}\, \Gc$ ($c$ is the speed of light), the bandwidth of  laser irradiation is equal to $100\, \MGc$, the natural linewidth $\gamma_{sp}=2\pi\times10^8\ \Hz$. We do not consider detuning of  laser wave from the $1S_{1/2} - 2P_{3/2}$ transition.
For these parameters from (\ref{eq18})  we get the value of saturation intensity  $I_0 \approx 45.6 \, \Wt/\ccm^2$.
At present time laser sources of continuous  Lyman-$\alpha$ irradiation can reach the maximal   power of $P=200\, \nWt$ \cite{Kolbe:2011}. For
the beam with the cross-sectional area $S \simeq 1.76\cdot 10^{-2}\, \ccm^2$  the  intensity value is
\begin{equation}
I=P/S=0.6 \cdot 10^{-3}\, \Wt/{\ccm}^2,
\end{equation}
so the ratio $I/I_0= 1.3 \cdot 10^{-5}$, which is not large enough for the fast laser cooling.

\subsection{Generation of continuous Lyman-$\alpha$ irradiation}
\label{subsec:17}

Production of the coherent  Lyman-$\alpha$ irradiation with $\lambda=121.56\ \nm$ is very difficult because
the non-linear crystals or tunable lasers are absent at this wavelength.
The well known $BBO$ crystal (barium borate) is the nearest to this optical region, however even $BBO$ crystal  is transparent  for the wavelengths $>190\ \nm$. In 1977   the four-wave resonant mixing in a gas was proposed for the Lyman-$\alpha$ irradiation.
The first Lyman-$\alpha$ sources were pulsed and employed the  Krypton gas \cite{Mahon:1978,Cotter:1979,Wallenstein:1980,Phillips:1993,Cabaret:1987,Marangos:1990,Batishe:1977}.

Usual Lyman  sources have a very small power because the generation of such irradiation by four-wave mixing is technologically difficult process.
The maximum reflectivity of mirrors is equal to $85\,\%$, the transmission coefficient  of lenses is $\sim 50\ \%$.
So the  arrangements producing this light should contain the minimal numbers of lenses and optical elements. Laser light should spread  in vacuum or $He$ to minimize  the intensity loss.

In  1999 K.S. Eikema proposed two  schemes using  the spectral transitions in mercury vapour for obtaining the cw Lyman-$\alpha$ emission \cite{Eikema:1999}.

\begin{figure}[htb]
\begin{center}
 \includegraphics[scale=0.58, angle=0]{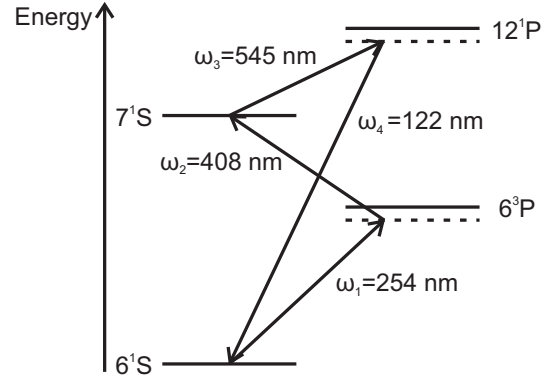}
\caption{Energy-level diagram of mercury levels used for generation of Lyman-$\alpha$ irradiation by sum-frequency four-wave mixing. Three fundamental beams at $254\ \nm$, $408\ \nm$ and $545\ \nm$ excite the atoms from $6^1 S$ state to $12^1 P$ followed by the spontaneous emission of Lyman-$\alpha$ irradiation.}
 \label{fig:FWM}
\end{center}
\end{figure}

The first scheme is shown in Figure \ref{fig:FWM} and based on resonant transitions between several atomic levels $6^1 S$, $7^1 S$ and $12^1 P$. They achieved the maximum power $3\ \nWt$ of Lyman-$\alpha$ emission at $121.9\ \nm $ using three fundamental beams.
 
 In the  second scheme  $6^{1} D$ and $6^{3} D$ levels were used  instead of $7^1 S$ level.
It turns out that the efficiency of the second method was twice  lower than  of the first one.
 Later the power of this Lyman-$\alpha$ source was improved till $20\ \nWt$ and spectroscopy measurements were performed on $1S-2P$ transition  in hydrogen  \cite{Eikema:2001}.  
 
The power of Lyman-$\alpha$ irradiation  is defined by the following equation

\begin{equation}
P_4=\frac{9}{4}\frac{\omega_1 \omega_2 \omega_3 \omega_4}{\pi^2 \epsilon^2_0 c^6}\frac{1}{b^2}\frac{|\chi^{(3)}|^2}{\triangle k^2} P_1 P_2 P_3 |G(b  \triangle k)|^2,
\label{eq26}
\end{equation}

where $\omega_1$, $\omega_2$, $\omega_3$ and $\omega_3$ are  the frequencies of the fundamental and generated beams correspondingly. $b$ is the confocal parameter, $\triangle k = k_4 - k_1 - k_2 - k_3$ is the phase mismatch, $N$ is the density of    the gas, $|\chi^{(3)}|$ is the third-order non-linear susceptibility. $P_1$, $P_2$, $P_3$ are the powers of the fundamental beams, $P_4$ is the power of Lyman   beam, $|G(b  \triangle k)|^2$ is the phase-matching integral, $\epsilon_0$
and $c$ are the fundamental physical constants. 

The third-order non-linear susceptibility of mercury factorizes \cite{Smith:1987}

\begin{equation}
\chi^{(3)}=\frac{N}{6\epsilon_0 \hslash^3 } \frac{\chi_{12} \chi_{34}}{\omega_{ng}-(\omega_1+\omega_2)}  ,
\label{eq27}
\end{equation}

where the partial susceptibilities are defined by the following way

\begin{equation}
\chi_{12}=\sum_m \left(\frac{p_{nm} p_{mg}}{\omega_{gm}-\omega_1}+ \frac{p_{nm} p_{mg}}{\omega_{gm} - \omega_2}\right)
\label{eq28}
\end{equation}
\begin{equation}
\chi_{34}=\sum_{\nu} \left(\frac{p_{n \nu} p_{\nu g}}{\omega_{g \nu}-\omega_4}+ \frac{p_{n \nu} p_{\nu g}}{\omega_{g \nu} + \omega_3}\right)
\label{eq29}
\end{equation}

 The susceptibility shows the enhancement at the small value of detuning $\delta=\omega_{gm} - \omega_2$.
 The phase-matching integral $|G(b  \triangle k)|^2$ takes its maximum value $46.3$ at $b \triangle k = -4$ \cite{Bjorklund:1975}.

For the Lyman-$\alpha$ irradiation phase-matching can be done by changing the temperature of mercury vapour or by adding another gas,  for example $Kr$. Usually the first way is chosen because the second one leads to the broadening of the mercury resonances and decrease of the Lyman-$\alpha$ yield.
From \ref{eq27} it follows that $P_4$ reaches its maximum value for $\triangle k = 0$. So one have to choose the optimum value of $\triangle k$, which depends on the experimental conditions.

The continuous irradiation is more preferable for laser cooling than pulsed one.
If the laser pulse duration is $t_p$,  the pulse repetition rate is $1/t_R$ and
$t_p \ll \tau \ll t_R$  then cooling efficiency will be in $t_p/t_R$ times lower than for the case of continuous laser radiation \cite{Lett:1988}. 
The cw laser irradiation has a smaller  linewidth than the pulsed irradiation, this reduces  the spurious optical pumping to the untrapped states during cooling process causing the loss of  particles from the magnetic trap.
The possible  solutions of the problem of optical pumping in antihydrogen atoms  were mentioned in \cite{Ertmer:1988}. It consists in use of circularly polarized light, a longitudinal magnetic field and   a second laser with another frequency to repump the atoms via the transitions $1S_{1/2}(m_F=-1/2) \rightarrow P_{3/2} \rightarrow 1S_{1/2}(m_F=-1/2)$, where $m_F$ is the projection of electron spin on the direction of the magnetic field \cite{Ertmer:1988}.

\begin{figure}[htb]
\begin{center}
 \includegraphics[scale=1, angle=0]{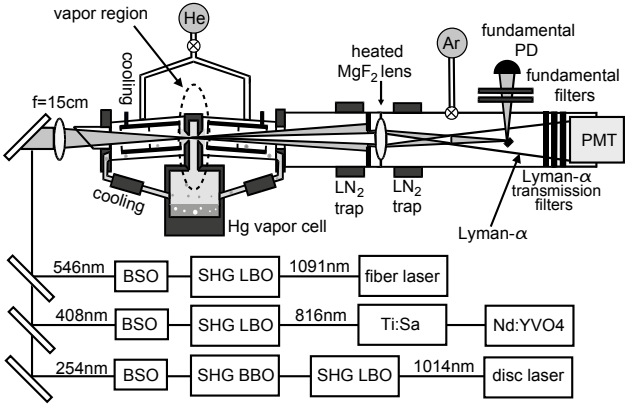}
\caption{The experimental setup used for the production  Lyman-$\alpha$  emission  \cite{Kolbe:2011}. Three fundamental beams from fiber laser, $Ti:Sa$ and disc lasers are frequency doubled in non-linear crystals ($LBO$ and $BBO$)  shaped by special optics ($BSO$) and focused in the mercury vapour cell, where the production of Lyman-$\alpha$ irradiation takes place. The separation of $121.56\ \nm$ light occurs due to the $MgF_2$ lenses and $122\ \nm$ filters. The Picture was taken from \cite{Kolbe:2012}.}
 \label{fig:Lyman2}
\end{center}
\end{figure}

In the following ten years the development of the Lyman sources has been in progress \cite{Walz:2001,Pahl:2005,Scheid:2009,Kolbe:2011}.
The  experimental setup  from \cite{Kolbe:2012} is presented in Figure \ref{fig:Lyman2}.
It has several advantages in comparison with the previous arrangements for the generation of Lyman-$\alpha$ emission \cite{Eikema:1999,Eikema:2001,Walz:2001}. One of the advantages  is a  more powerful source of $1091\ \nm$.
The resultant  emission power  could be increased  due to the increasing of the power of  each wave in the scheme.

The   $Yb:YAG$ disc laser  generates  emission at $1014\ \nm$. 
Then the beam passes through a lithium triborate  crystal ($LBO$) producing $507\ \nm$ emission and after that through $\beta$-barium borate crystal ($BBO$) to obtain  the $UV$ light at $254\ \nm$. The   power of the $UV$ light may reach $750\ \mW$ if the   disc laser has its maximum power $4.9\ \W$. At such high power $BBO$ crystal is subjected to degradation so the work power is limited to $2\ \W$.  

The second beam at $408\ \nm$ is produced by the following way. 
$Ti:Sa$ laser pumped by $Nd:YVO_4$ laser gives the $816\ \nm$ irradiation with the power  $1.6\ \W$.
From near-infrared irradiation at $816\ $ after passing the $LBO$ crystal we get   $500\ \mW$ of blue light at $408\ \nm$ .
The work output power equals $300\ \mW$. 
 
  The third beam  at $546\ \nm$ is obtained with a fiber laser system with $10\ \W$ at $1091\ \nm$ and $LBO$ crystal. The maximum power of the third beam can reach $4\ \W$, however to forbid the spontaneous damage of the $LBO$ crystal  the fiber laser 
  works at $740\ \mW$, producing the $280\ \mW$ of green light.
  
\begin{figure}[htb]
\begin{center}
 \includegraphics[scale=0.33, angle=-90]{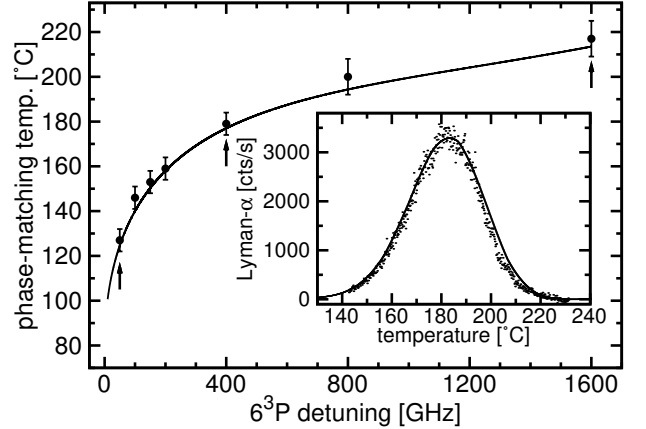}
\caption{The   phase-matching temperature versus the $6^1S-6^3P$ detuning. The points correspond to the maximal Lyman-$\alpha$yield for various values of the detuning. Inset: the Lyman-$\alpha$ yield as a function of cell temperature at the detuning $400$ Gz. Points denote experimental line, solid line is the fit  \cite{Kolbe:2012}. }
 \label{fig:Kolbe3}
\end{center}
\end{figure}

The beams are shaped by pairs of spherical and cylindrical lenses,
overlapped with dichroic mirrors and focused by fused  silica lens with focal length of $15\ \cm$  into the mercury cell, where the four-wave frequency mixing  takes place. The pressure in the mercury cell is supported at $10^{-7}$ mbar to prevent the absorption of $UV$ light. The mercury cell can be heated till $240\  \Cc$ and the density of atoms is $1.1 \times 10^{24}\ \m^{-3}$.
The Lyman-$\alpha$ yield as a function of mercury temperature is shown in Figure \ref{fig:Kolbe3}, the phase-matching temperature as a function of $6^1S-6^3P$ detuning is also depicted for maximum Lyman-$\alpha$ yield \cite{Kolbe:2012}.

 The cooling outside the focus region is necessary to avoid the condensation of the mercury on the optics.
 Separating the Lyman-$\alpha$ emission from the fundamental beams is  realized due to the $MgF_2$ lens.
 The focal lengths differ for various wavelengths. The tiny mirror behind the cell reflects the fundamental beams to the side, however it cast a shadow to the Lyman beam and produce $30 \%$ of the losses.
 Then the light passes through $121.56\ \nm$ filters.
 There is a photomultiplier tube ($PMT$) used for the detection of Lyman photons behind the  filters.
 The beam of $254\ \nm$ wavelength is tuned close to  $6^1S$ - $6^3P$ resonance, the blue laser ($408\ \nm$) establishes
a two photon resonance between $6^1 P$ and $7^1S$ state.
The third beam ($545\ \nm$) is in resonance with the transition $7^1S$ - $12^1 P$.
The sum-frequency gives Lyman-$\alpha$ wavelength at the transition $12^1P$ - $6^1 S$.

At the maximal power of the fundamental beams $750\ \mW$, $500\ \mW$ and $4\ \W$ described system can produce Lyman-$\alpha$ irradiation with the total power equal to $140\ \nWt$. 
The increase of the yield of Lyman  emission seems feasible due to increase of the transmission of the lenses and the power of the fundamental beams.

Nevertheless the generation of emission of Lyman-$\alpha$ wavelength  is technologically complicated and the development of such lasers is a difficult task. An alternative way can consist of the use of other spectral transitions in (anti)hydrogen or other cooling methods.

\section{Other cooling schemes for $\bar{H}$ }
\label{sec:18}

\subsection{Two-photon pulsed laser cooling}
\label{subsec:19}

 \ \ \ Theoretically it is possible to utilize  other spectral 
 transitions for  laser cooling of antihydrogen atoms.
However   experimental attempts to do this were not undertaken.
In \cite{Zehnle:2001} the authors discuss  cw  Doppler cooling of hydrogen atoms at two photon   $1S - 2S$ transition. The state $2S$ itself is not useful for laser cooling,
because its lifetime is large $1/8\, \cc$ enough, but the idea was  to couple the $2S$ to the $2P$ state by a microwave radiation and the resulting state would have a much shorter lifetime.
The numerical simulations show that this method allows to get the temperatures, which correspond  to the recoil energy of antihydrogen for the $1S\rightarrow 2S$ transition, the Lamb shift between $2S$ and $2P$ is small and has not been taken into acoount. For the intensity $I\sim 10^3\, \Wt/\cm^2$ the scheme leads to $18\%$  ionization losses per fluorescence cycle.

\begin{figure}[htb]
\begin{center}
 \includegraphics[scale=0.37, angle=0]{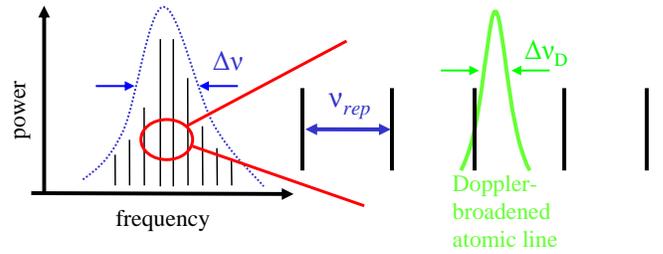}
\caption{Left: Power of the spectrum lines of the ultrafast pulse versus the frequency of the line. Right: Several lines of the pulse spectrum and the Doppler-broadening atomic transition of width $\triangle \nu_D$ \cite{Kielpinski:2006}.}
 \label{fig:ultrafast}
\end{center}
\end{figure}

It was proposed to use the ultrafast pulsed radiation \cite{Kielpinski:2006}
instead of the continuous one at $243\, \nm$.
The ultrafast pulsed radiation from the mode-locked laser has several advantages.
It is a high spectral resolution and scaterring rate larger by a factor of $\sim 10^4$ than its value for the cw case.
The peak power of such lasers are also higher than the power of continuous wave lasers, for example pulsed $Ti:Sa$ laser has a power up to $1\ \W$. The restriction for the maximum used power is defined by one-photon ionization from excited state   leading to losses of (anti)atoms from the magnetic trap.
The scaterring rate defines the efficiency of laser cooling, its high value gives a much faster cooling.

The pulse of the ultrafast laser is a comb of sharp lines, the frequency of the line is $\nu_k=\nu_0+k\nu_{rep}$, where $k$ is the integer number, $k_0$ is the optical carrier frequency, $\nu_{rep}$ is the repetition rate. The velocity-selective scaterring takes place if one of the comb lines is resonant with the atomic transition as it is shown in Figure \ref{fig:ultrafast} \cite{Kielpinski:2006}. If the Doppler width of the line is $\Gamma_D$ and $\nu_{rep}\gg \Gamma_D$ then
scattering rates on the other spectrum lines   are reduced in $(\Gamma/\nu_{rep})^2$ times and the near-resonant line of the comb gives  the dominant contribution. As a result the continuous wave laser  cooling is less effective than two-photon cooling, because only one line of the comb in resonance with the atomic transition. During the two-photon process all the comb lines give significant contribution to the scattering because the transition pathways of the photons add coherently.   

\begin{figure}[htb]
\begin{center}
  \includegraphics[scale=0.48, angle=0]{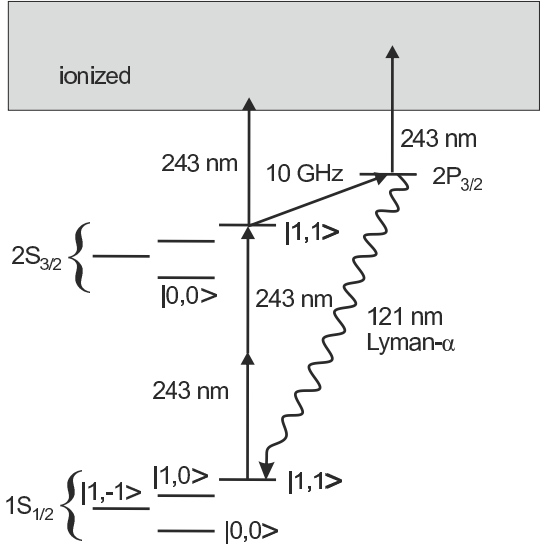}
\caption{Excitation scheme for   laser cooling of magnetically trapped (anti)hydrogen. The two-photon pulsed laser irradiation at $243\ \nm$ excites the atoms from the ground $1S_{1/2}\vert 1,1\rangle$ state to the $2S_{1/2}\vert 1,1\rangle$ state. Microwave irradiation at $10$ GHz quenches $2S_{1/2}\vert 1,1\rangle$  state to the $2P_{3/2} \vert 2,2 \rangle$ state. Atoms irradiate spontaneously Lyman-$\alpha$light and come back to the ground state $1S_{1/2}$, $F$ is the magnetic quantum number, $m_F$ is its projection  \cite{Kielpinski:2006}.}
 \label{fig:4}
\end{center}
\end{figure}

The possible excitation  scheme for the laser cooling of (anti)hydrogen in the magnetic  trap is depicted in Figure \ref{fig:4} \cite{Kielpinski:2006}.
The light from the pulsed laser source is tuned to the $1S-2S$ spectral transition for antihydrogen atom.
The meastable $2S_{1/2}$ with $m_F=1/2$ state quenches to  the $2P_{3/2}$ state with $m_F=3/2$ by microwave irradiation with frequency $f_{\micr}$  and circular polarization $\sigma^{+}$.
The splitting of the (anti)hydrogen sublevels are very sensitive to the value of the trap magnetic field, so 
the  frequency of the microwave irradiation have to be chosen appropriately.
 For  zero magnetic field it would be equal to $10$ GHz, at $1\ T$ the frequency $\sim 20$ GHz. So the light excites the atoms from the ground state $1S_{1/2}$  with $m_F=1/2$ to  some stretched state which is a superposition of $2S_{1/2}$ and $2P_{3/2}$ states and can survive in the magnetic trap.
Then atoms irradiate Lyman-$\alpha$ photon and come back to the ground state.

For efficient two-photon process   on $1S-2S$ transition we   need powerful  laser source,  the upper limit for the intensity of the pulsed laser source comes from the one-photon ionization rate from the excited state by $243\ $nm irradiation. Nevertheless
we think that this scheme may be a good alternative to Lyman-$\alpha$laser cooling of antihydrogen to milliKelvin temperature range. Numerical simulations showed that for the power $1.6\ \W$ of the quenching irradiation and $60\ \kWt/\cm^{-2}$ the ionization losses consist $ 5 \%$, i.e. they are less than for cw case.
\begin{figure}[htb]
\begin{center}
 \includegraphics[scale=0.49, angle=0]{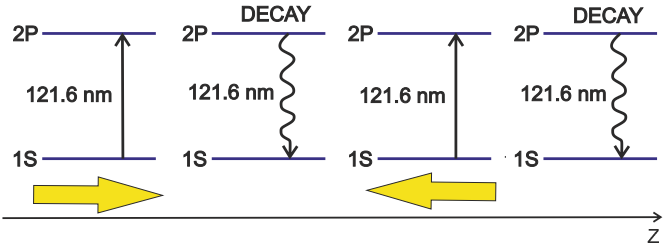}
\caption{One cooling cycle, consisting of a $1S-2P$ transition due to an application of the $\pi$ pulse  with negative frequency detuning $\delta$ in the positive direction  and a spontaneous emission $2P-1S$, the use of the  $\pi$ pulse in the negative direction and the same transitions.}
 \label{fig:5}
\end{center}
\end{figure}

\subsection{Laser cooling by  $\pi$ pulses}
\label{subsec:20}

 Radiation cooling by $\pi$ pulses was suggested on the hydrogen Lyman-$\alpha$ transition in \cite{Palmer:1986}.
The train of coherent pulses in the opposite directions irradiate the cooling volume. The pulse train is short in comparison with the radiative lifetime of the transition. The photon from the wave moving from the left is absorbed, exciting the atom to the upper $2P$ level, followed by the spontaneous emission to the ground state $1S$, see Figure \ref{fig:5}.
To retard the motion this pulse should be detuned above the resonance.
Then the pulse from the wave moving from the right is absorbed, the wave frequency is shifted below the $1S - 2P$ transition.
During the cooling a spontaneous emission does not occur. To escape the phase disparity the specific zero-field period is necessary between $\pi$ pulses. The numerical analysis of rate equations showed that the rate of this  cooling method is larger than for  Doppler cooling, but it can be reached at the expense of considerably higher average power density.

\begin{figure}[htb]
\begin{center}
 \includegraphics[scale=0.49, angle=0]{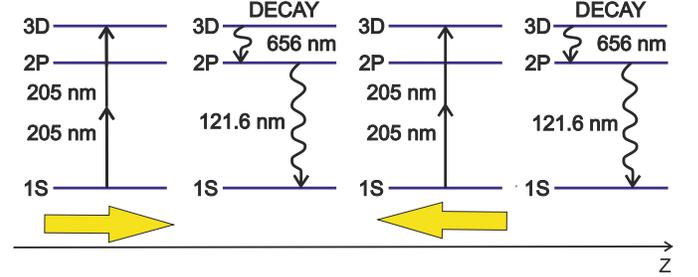}
\caption{One cooling cycle consisting of a $1S-3D$ transition due to the application of the  $\pi$ pulse  with negative frequency detuning $\delta$ in the positive direction  and  spontaneous $3D-2P$ and $2P-1S$ emissions, the use of the  $\pi$ pulse in the negative direction and the same transitions.}
 \label{fig:55}
\end{center}
\end{figure}

Another scheme for laser cooling of H with $\pi$ pulses  was suggested in \cite{Allegrini:1993}.
The scheme utilizes the pulsed radiation at a wavelength longer than the Lyman-$\alpha$ radiation.
Powerful laser generates   $\pi$ pulses at the wavelength  $205\ \nm$ and establishes  the two-photon resonance with the transition $1S-3D$. Then the excited atom come  back to the ground state emitting the photons at $656\ \nm$ and $121.56\ \nm$ wavelengths.
Simulations showed that the final temperature achieved with a monochromatic light  consisting of pulses with the duration $10$ ns and the power $100\ \kWt$  is close to the recoil temperature for Lyman-$\alpha$ transition of hydrogen $ \sim 0.55\ \mK$. The limiting factors in this scheme are the very low diffusion coefficient on the $3D-2P$ transition and the recoil limit on the $2P-1S$ transition. Because of the spurious ringing effects this method might be applied only as a final step to $H$ atoms precooled via another method.

However all these methods  suffer  from  ionization losses which should be mitigated when we  deal with  small amounts of antiatoms in a magnetic trap. The first attempt to do this has been made in \cite{Wu:2011}.
The scheme supposes the use of nanosecond-pulsed  laser at  $243\ \nm$ for $1S - 2S$ two photon transition  and continuous wave laser at $656\ \nm$ for $2S - 3P$ transition in antihydrogen. The atoms are excited selectively depending on their velocity, which is defined by spatial atom position in an optical lattice created by cw laser waves. The scheme presents a combination of pulsed Doppler  and Sisyphus cooling schemes. The minimal temperature  is $1.8\ \mK$. Spin-flip transitions from the $3P$ excited state are suppressed in a high magnetic field.

\subsection{Boofer cooling of (anti)hydrogen}
\label{sec:21}

The large recoil energy of the (anti)hydrogen   impedes  reaching   temperatures below $\m\K$ in laser cooling.  
 Evaporative cooling leads to the loss of particles and progresses slowly,  because the cross section of elastic scattering between hydrogen atoms has a  small value. 
  
Adding impurity atoms with a large cross section in a cooled gas can lead to a more rapid energy exchange  between   hot and cold atoms and serve as a source of thermalization. Laser is used for cooling of the impurity atoms which collide with the main gas and take away their energy access due to these collisions. 

 Thermalization rate  depends on the cross section of the colliding atoms, their masses and densities.  
 The energy exchange  goes faster for the lighter atoms of the buffer gas. 
 Boofer cooling is not dependent on internal states of the atoms. It works for the all kinds of paramagnetic species including molecules.  The boofer and cooled gases has to be trapped to isolate   the system from the walls of a vacuum chamber.
 At first the technics of boofer cooling for neutral atoms was described in 1995 \cite{Doyle:1995} and performed in 1999 \cite{deCarvalho:1999}.
 
 The  boofer cooling of hydrogen was proposed in \cite{Derevianko:2001,Cote:2000}.
 This idea belongs to Professor D. Kleppner and Professor J.M. Doyle. 
 To prevent the coalescence of different atoms and their loss from the trap the scaterring length has to be positive value.
   It   is satisfied if the collisions  occur  mainly in the triplet channel ie for spin-polarized atoms \cite{Cote:2000}.
The atomic cloud with a given polarization can be created in a magnetic trap with a fairly smooth magnetic field.
The shape of the singlet and triplet potentials are discussed in \cite{Cote:2000}. Singlet potential has a deep minimum and triplet potential  may   have not such clear minimum.

Therefore the interaction can occur both in the singlet and triplet channels. 
When the spins of colliding atoms are parallel the atoms interact in the triplet channel, antiparallel - in the singlet one.
Generally for not polarized spins the collisions occur in the singlet state. It can  result to the coalesence of atoms, formation of bound states,  spin flips, which in turn leads to a loss  of atoms from the magnetic trap.

The method of buffer cooling suggested in \cite{Derevianko:2001,Cote:2000} is as follows.
A small amount of   cold alkali-metal atoms is  mixed with hydrogen atoms in a magnetic trap. The
 boofer gas   can be presented $^{23}Na$, $^{39}K$, $^{87}Rb$, $^{133}Cs$ atoms or $^7Li$ isotopes.  At   temperatures considered the scaterring takes place in s-wave. Fermions  $^6Li$ cannot be used as boofer gas for cooling because  the s-wave scaterring is forbidden for them.

For energies   $\lesssim 3\ \m\K$  the  cross section   $\sigma=8 \pi a_{H-H}^2$ for hydrogen-hydrogen and $\sigma=4 \pi a_{A-H}^2$ for alkali-atom-hydrogen collisions,  where $a_{H-H}$ and $a_{A-H}$ are  the corresponding  scaterring lengths.
Triplet cross section for $^7Li-H$ case is approximately 1467 larger  than for hydrogen atoms. 
The ultracold cross section for the  $Rb-H$ purely spin-polarized pairs is 860 times larger than for $H-H$ interaction.
The cross section between $Na$ and $H$ atoms is 640 times larger than for the pair of hydrogen atoms.
The interaction potential and corresponding scaterring lengths  in singlet and triplet channels between various atoms and hydrogen are known and were calculated in \cite{Cote:2000} by various methods.

The large cross section between alkali-metal and hydrogen atoms might  lead   to a  more effective evaporative cooling. Evaporative cooling can be applied for  obtaining of Bose-Einstein condensates.
To separate hydrogen from alkali-metal atoms    one can use the  microwave irradiation   which produces spin flips of boofer atoms. 

In the limit of low temperatures the triplet scattering length for hydrogen atoms is  $a(H-H)=1.2\ \bohr$, for lithium and hydrogen atoms    $a(^7Li-H)=65\ \bohr$. For lithium atoms the scattering length is large and negative   $a(^7Li-^7Li)=-27.6\ \bohr$, thefore the lithium atoms coalesce with each other. Such behaviour presents a problem because 
  one can not make an experiment at large lithium concentrations. The choice of the most suitable experimental parameters, at which the cooling  proceeds effectively, can be very important. The numerical simulations of particles dynamics are relevant.  
 Therefore a small admixture of lithium atoms can  speed up the process of cooling hydrogen, but their high density probably leads to a loss of particles   out of the trap.

It would seem the  technique of buffer  cooling does not work with antimatter due to annihilation.
 Nevertheless P.K. Sinha and A.S. Ghosh \cite{Sinha:2005}   explored the interaction  dynamics for $\bar{H}-Li$ system  at temperatures  from  $\pK$ to $32\ \K$   and  conclude that the 
  $^7 Li$ gas can be used for boofer cooling of atomic antihydrogen also. 
 The sign of the interaction potential between antihydrogen and lithium depends on the temperature and on the quantum number of an antihydrogen state.
For the antihydrogen in the ground state  the potential is repulsive but for $n>6$  it stays attractive from $0$ to $3\ \K$ \cite{Sinha:2005}. Higher excited states has no qualitative effect on the behaviour of interaction potential, however lead to an  increase of the cross section and shift of the potential curve to the region of higher energies.  
 Lithium has a large dypole polarizability in comparison with antihydrogen atoms, so the van der  Waals  force mainly   determines the low energy behaviour.

Recently scientists   suggested  a method to escape annihilation. $GBAR$ collaboration plans to synthesize the positively charged ions of antihydrogen, mix them with  positive charged beryllium ions, cooled by laser irradiation.
The Coulomb interaction will impede annihilation. It is planed to achieve the temperatures $\sim 20\ \mu \K$.
Unfortunately the output of antihydrogen ions is several orders less than  antihydrogen atoms.

\subsection{Sympathetic cooling of antihydrogen ions}
\label{subsec:22}

$GBAR$ collaboration  at $CERN$ \cite{Perez:2008,Perez:2011} suggested not to cool the antihydrogen but  obtain it in a very cold state.
In their scheme   antihydrogen is obtained due to the photodetachment of positron from  antihydrogen ion 
$\bar{H}^+$ by the laser irradiation at the wavelength  $313\ \nm$. Antihydrogen ions is produced through the reaction of   proton with Rydberg positronium  
\begin{equation}
\bar{p}+Ps^* \rightarrow \bar{H} + e^-
\end{equation}
and the reaction of antihydrogen with the positronium
 \begin{equation}
\bar{H}+Ps^* \rightarrow \bar{H}^+ + e^-.
\end{equation}
However the yield of  antihydrogen ions in the ground state is very small.
It was proposed to excite the positronium to the $3D$ level in the first reaction. It leads to the increasing of  $1S$ $\bar{H}$ production by a factor of 3 \cite{Mitroy:1994,Mitroy:1995} in comparison with the  positronium in  $2P$ state.

The antihydrogen ions will be captured and decelerated to $10$ eV.
 Then they will be  sympathetically cooled by   $Be^+$ ions.
  At first berillium ions in turn will be subjected to the Doppler laser cooling.
 The ions of alkali-earth elements  could be easily cooled to $\mu \K$ temperatures. 
 $Be^+$ ions is the most light alkali-earth metal with Doppler limit equal to $0.5\ \mK$. 
 At this temperature   Coulomb repulsion is strong enough in comparison with their kinetic energy and beryllium ions are organized in Wigner crystal.
   After that  $\bar{H}^+$  ions will be transferred to the precision trap for further sympathetic cooling, where berillium ions will be cooled by Raman sideband cooling technique to the neV energy.  
  
 To know the cooling time of $\bar{H}^+$ is very important because of possible positron photodetachment  by the   laser light. The lifetime of $\bar{H}^+$ is less than second under conditions of the $GBAR$ experiment \cite{Hilico:2014}.   The computer simulations \cite{Hilico:2014} has shown that the efficient and fast sympathetic Doppler cooling of energetic antihydrogen ions is possible in the presence of  $HD^+$ ions. The cooling rate depends on the mass ratio and  $HD^+$ ions act as a conduit between beryllium and antihydrogen ions.
  Raman cooling is   possible within the time less than one second \cite{Hilico:2014}.
  
\begin{figure}[htb]
\begin{center}
 \includegraphics[scale=0.48, angle=0]{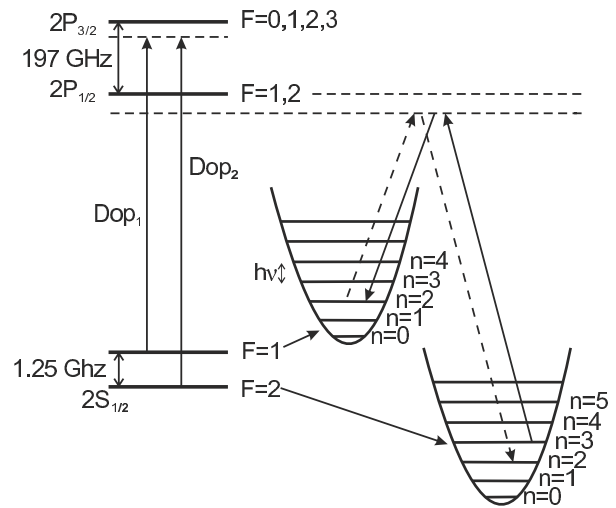}
\caption{Energy levels of $Be^+$ including the harmonic oscillator energy at the ground levels in the magnetic trap. The solid arrows indicate the stimulated Raman transition and the dashed arrows the spontaneous Raman transition. $Dop_1$ and $Dop_2$ are the Doppler cooling and repumping transitions \cite{Hilico:2014}.}
 \label{fig:6}
\end{center}
\end{figure}

  The Figure \ref{fig:6} shows the electronic $Be^+$  levels and illustrates the idea of Raman sideband cooling. Doppler   and repumping transitions are also indicated.
  
The idea of the Raman sideband cooling consists in a decrease of the vibrational number of the ground state atom, when it is in the potential of the magneto-optical trap \cite{Dehmelt:1975,Diedrich:1989}. Cooling is possible if the laser linewidth and the recoil energy is smaller than the separation of the atomic energy levels. Counter-propagating laser $\pi$ pulses induce the Raman transitions between 
$\vert F=2,n\rangle$ and $\vert F=1,n-1\rangle$ states followed by spontaneous Raman transitions from $\vert F=1,n-1\rangle$ to $\vert F=1,n-2\rangle$. Repeating this procedure one can prepare $Be^+$ ions in the fundamental state. Elastic interactions with antihydrogen ions will transfer the kinetic energy from $\bar{H}^+$ to laser cooled beryllium ions.

 Negative osmium ions has a bound electric-dipole transition and they can be laser cooled to the sub-mK temperature range.  The sympathetic cooling of the antiprotons  by osmium ions was suggested in \cite{Widmann:2005}. These cold antiprotons could be used for the synthesis of cold antihydrogen by resonant charge-exchange with Rydberg positronium.

\section{Collimated   atomic beams}
\label{subsec:23}

Several experiments at $CERN$ use the beam of spin-polarized  antihydrogen atoms formed in the cusp trap.
$FLAIR$ collaboration in $GSI$  will utilize such beam not only for precision spectroscopy measurements  but also for experiments 
on study of antihydrogen collisions with other particles.
In particular, a great interest presents an exploration of interaction between atomic  antihydrogen and hydrogen, molecules and heavy atoms from both   theoretical and practical point of view. These investigations could give signatures of annihilation  between antimatter and matter in a cosmic space.

$FLAIR$ will study the interaction potential between atomic antihydrogen and   atoms and molecules, annihilation processes and formation of positronium and antiprotonic atoms, search for scattering resonances and excitation exchange between 
 $\bar{H}^*$ and $H$. 
 Interaction between excited  $\bar{H}^*$ and matter has a practical importance because antihydrogen is formed in a highly excited states, it hinders   cooling and precision spectroscopy of $\bar{H}$,  a study of antihydrogen dynamics in such mixtures is important. 
 
The interaction  with positronium target is of great interest because it leads to the formation of $\bar{H}^+$  and opens up the new possibilities in production of antihydrogen molecules due to associative detachment.
\begin{figure}[htb]
\begin{center}
 \includegraphics[scale=0.5, angle=0]{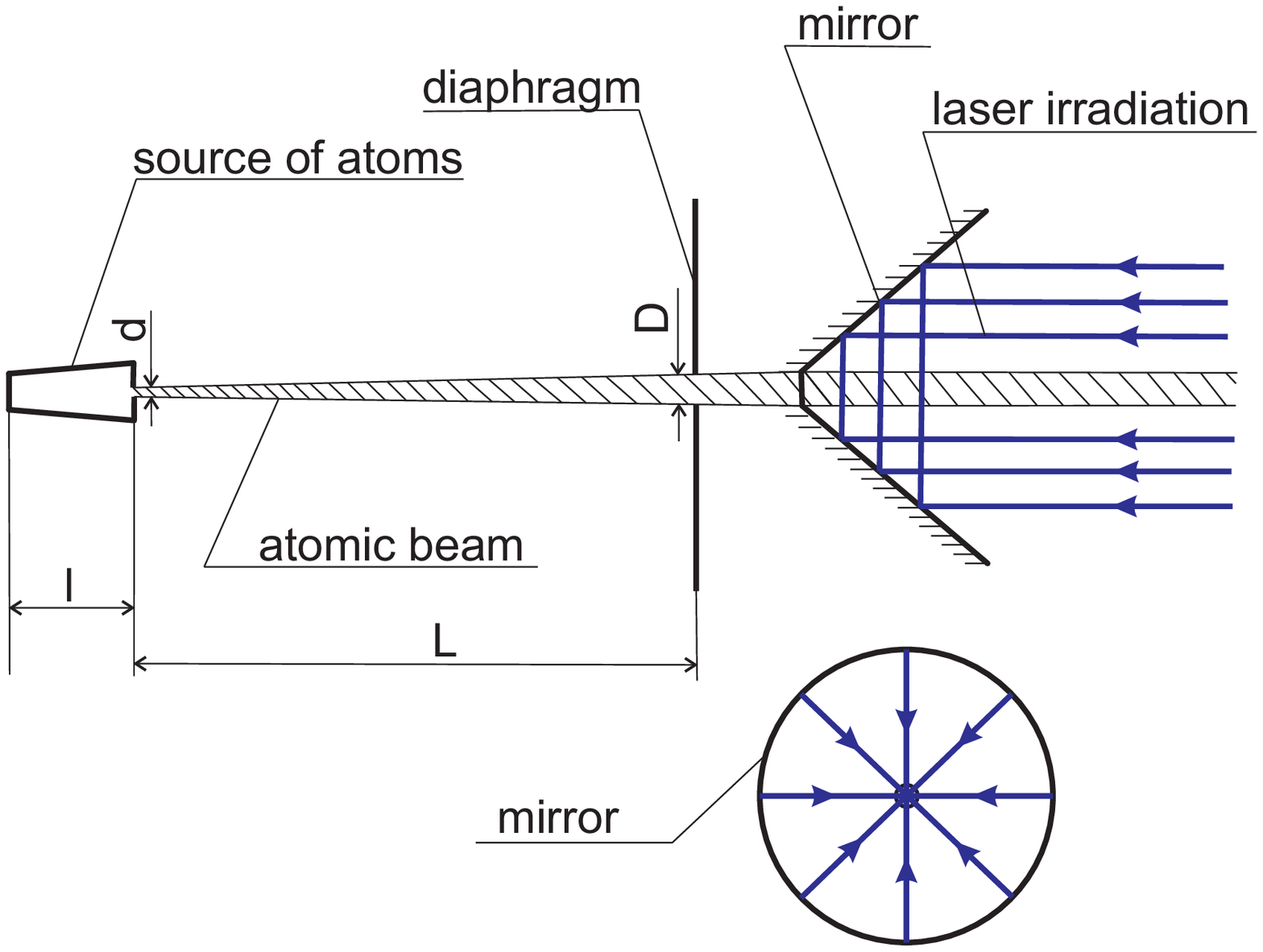}
\caption{The scheme of radiational collimation of an  atomic beam   due to the Doppler colling  in the perpendicular direction to the axis of the beam.}
 \label{fig:collimation}
\end{center}
\end{figure}
 In all these experiments it is very important to have a collimated beam of high quality. 
Theoretical  and experimental investigations   showed that such collimation could be done by radiation pressure of light irradiation \cite{Balykin:1984,Balykin:1985,Letokhov:1977}. 

  In Fig.\ref{fig:collimation} the   radiational collimation scheme of an atomic beam is shown. This   scheme  has been used  for the experiment with sodium atoms \cite{Balykin:1984}. The atomic beam from the orifice of the source  pass through the diaphragm and  moves to a region of space  illuminated by  an axially-symmetrical light field.  The axially-symmetrical laser field  due to mirror reflection of laser irradiation cools the atoms in the perpendicular to the beam direction. The surface of mirror is a cone with generatrix which forms   angle of 45 degrees with respect to  the beam axis. 
 Therefore   the maximum intensity is  achieved on the axis of the atomic beam and   radiation pressure force is directed to the beam axis.
  
Beam collimation or cooling of (anti)atoms in the transverse direction is carried out due to the Doppler cooling mechanism described in Section \ref{subsec:4}. 
For example in experiment with sodium atoms the temperature, corresponding to the transverse motion of atoms, decreased   from $42$ mK to $3.5$ mK during  $3\times 10^{-5}$ s when the power on the beam axis was equal to   $P=60$ mWt. 
The resonant transition  $3S-3P$ was used for Doppler cooling. 

Usually atoms effuse from the owen at some temperature $T$. If the Knudsen number $K= {\lambda} / l\lesssim 1$, where $l$ is the length of the atomic source, $ {\lambda}$  is the mean free path of the atom, then the distribution of atoms is Maxwellian, so atoms are moving from the source with the most propable velocity $v=\sqrt{2k_B T/m}$. 
For the scheme at Fig.\ref{fig:collimation} the  divergence angle of the beam equals to $\alpha=(D-d)/(2L)$, where $d$ is the diameter of the orifice, $D$ is the size of the orifice in the diaphragm, $L$ is the distance between diaphragm and the orifice of the atomic source.  The angle $\alpha$ is small for $d,D\ll L, \ D\ll L$. Further collimation of the atomic beam by   laser irradiation has to diminish  the divergence of the beam. 
 
The minimal temperature corresponding to a stationary distribution of   transverse velocities in the region of collimation
\begin{equation}
T=\frac{\hslash \gamma}{2k_B} \left(\frac{\gamma}{|\delta \omega|} + \frac{|\delta \omega|}{\gamma} \right)
\end{equation}
  and the corresponding angle of  the beam collimation 
  \begin{equation}
 \theta = \frac{\sqrt{2 k_B T/m}}{\bar{v}_z}
 \end{equation}
where $\bar{v}_z$ is the average velocity of atoms along the axis of the atomic beam,   $m$ is the atomic mass, $\delta \omega=\omega_l-\omega_0$ is the laser frequency detuning from the resonance atomic frequency $\omega_0$ \cite{Balykin:1985}.

The laser collimation of atomic beam is usefull for spectroscopy because it can reduce the first-order and second-order Doppler shifts on the orders of magnitude. Laser light can also be used for the spatial separation of the atoms in various internal states.

\section{Conclusions}
\label{sec:24}

Laser cooling is of great importance for antihydrogen research.
A large number of experiments devoted to study of antihydrogen  and antimatter are already underway in $CERN$
and planned in $GSI$ research facility.
Among their goals are verification of basic laws of nature, such as $CPT$ symmetry, equivalence principle, study of   new aspects of nuclear physics etc. 
Without a laser cooling of $\bar{H}$ it is impossible to achieve a high accuracy in these experiments.
Today the relative accuracy of $GS-HFS$ splitting of $H$ ground state in $10^{10}$ times higher than for antihydrogen, the electrical neutrality of atomic hydrogen is confirmed with  $10^{-21}$ precision, by $10^{13}$ orders of magnitude better than for antihydrogen. The measurement of $1S-2S$ transition in $\bar{H}$ hasn't been made at all.

In magnetic trap used at $CERN$ the temperature of atomic $\bar{H}$ is approximately equal to  $50\ \mK$. To improve the accuracy of the mentioned spectroscopic measurements  this temperature  should be in $10$ times lower.
The gravitational experiments need temperatures of $\mu \K$ range. 
In the magnetic trap the lowest hydrogen temperatures  $\sim 3-11\ \mK$ were achieved only by means of laser cooling.
 Laser can be used for the collimation of the antiatomic beam. The Bose-Einstein condensates cannot  be obtained without  laser cooling technics. The study of quantum properties both for hydrogen and antihydrogen is not possible without laser cooling.
 
The  $1S_{1/2} - 2P_{3/2}$  spectral  transition in (anti)hydrogen atom is the most suitable for the laser cooling. The corresponding  laser sources  of  the wavelength $\lambda=121.6\ \nm$ already exist, but their power is not sufficient for the fast and efficient cooling. The maximal power of continuous Lyman-$\alpha$ sources is $200\ \nWt$, the power of pulser sources $1\ \mWt$. The creation of powerfull Lyman-$\alpha$ sources is very important complicated technical task. The continuous sources at this wavelength could be used for experiments with magneto-optical trap.

Laser cooling has several advantages over other methods such as evaporative cooling, sympathetic cooling  and cooling due to a contact with cold walls of  experimental setup.
 Evaporative cooling leads to a loss of atoms from the trap. Sympathetic cooling of hydrogen is difficult to implement in practice, however it seems promising and possibly can be applied for antihydrogen. 
 The cooling owing  to a contact with the cold  walls of vacuum chamber  also leads to the undesired   recombination with the surfuce and loss of  atoms. 
 The alternative way can be a use of other transitions in the antihydrogen atom  or  new  laser cooling schemes.  Several attempts in this direction have been made already.

The great experience of cooling of the neutral atoms exists, the physical mechanisms of laser cooling have been studied.
The experimental results are in agreement with the known theoretical models. Many atom's species have been laser cooled, the     production of   (anti)hydrogen at $\sim \mK$ temperature has to be the next step. Cold antihydrogen has a great importance for fundamental science. 
The hydrogen $1S-2S$ transition is a promising candidate for an optical clock. The search of other possible applications of (anti)hydrogen is also significant important.

\section{Acknowledgments} 

The work was carried out with the financial support of FAIR-Russia Research Centre.
The authors are grateful to A.V.Masalov and N.N. Kolachevsky for the useful discussions and comments.


%

%
% BibTeX users please use
% \bibliographystyle{}
% \bibliography{}

\end{document}